\documentclass[fleqn,usenatbib]{mnras}

\usepackage{newtxtext,newtxmath}

\usepackage[T1]{fontenc}

\DeclareRobustCommand{\VAN}[3]{#2}
\let\VANthebibliography\thebibliography
\def\thebibliography{\DeclareRobustCommand{\VAN}[3]{##3}\VANthebibliography}

\usepackage{graphicx}	
\usepackage{amsmath}	
\usepackage{amssymb}	
\usepackage{siunitx}    

\title[WHIM Tracers at UV-wavelengths]{
Tracing the 10$^7$ K Warm-Hot Intergalactic Medium with UV absorption lines
}

\author[A. Yrupe Fresco et al.]{
A. Y. Fresco,$^{1}$\thanks{E-mail: yrufresquito@gmail.com (AYF)}
C. P\'eroux,$^{2,3}$
A. Merloni,$^{1}$
A. Hamanowicz,$^{2}$
and R. Szakacs $^{2}$
\\
$^{1}$Max-Planck-Institut f\"ur Extraterrestrische Physik (MPE), Giessenbachstrasse 1, D--85748 Garching, Germany\\
$^{2}$European Southern Observatory, Karl-Schwarzschildstrasse 2, D-85748 Garching bei M{\"u}nchen, Germany\\
$^{3}$Aix Marseille Universit\'e, CNRS, LAM (Laboratoire d'Astrophysique de Marseille) UMR 7326, 13388, Marseille, France \\
}

\date{Accepted XXX. Received YYY; in original form ZZZ}

\pubyear{2015}

\begin{document}
\label{firstpage}
\pagerange{\pageref{firstpage}--\pageref{lastpage}}
\maketitle

\begin{abstract}
Today, the majority of the cosmic baryons in the Universe are not observed directly, leading to an issue of "missing baryons" at low redshift. Cosmological hydrodynamical simulations have indicated that a significant portion of them will be converted into the so-called Warm-Hot Intergalactic Medium (WHIM), with gas temperature ranging between 10$^5$--10$^7$K. While the cooler phase of this gas has been observed using O\,VI and Ne\,VIII absorbers at UV wavelengths, the hotter fraction detection relies mostly on observations of O\,VII and O\,VIII at X-ray wavelengths. Here, we target the forbidden line of [Fe\,XXI] $\lambda$ 1354\AA\ which traces 10$^7$K gas at UV wavelengths, using more than one hundred high-spectral resolution  (R$\sim$49,000) and high signal to noise VLT/UVES quasar spectra, corresponding to over 600 hrs of VLT time observations. A stack of these at the position of known DLAs lead to a 5-$\sigma$ limit of $\mathrm{log[N([Fe\,XXI])]<}$17.4 (${EW_{rest}<22}$m\AA), three orders of magnitude higher than the expected column density of the WHIM $\mathrm{log[N([Fe\,XXI])]<}$14.5. This work proposes an alternative to X-ray detected 10$^7$K WHIM tracers, by targeting faint lines at UV wavelengths from the ground benefiting from higher instrumental throughput, enhanced spectral resolution, longer exposure times and increased number of targets. The number of quasar spectra required to reach this theoretical column density with future facilities including 4MOST, ELT/HIRES, MSE and the Spectroscopic Telescope appears challenging at present.  Probing the missing baryons is essential to constrain the accretion and feedback processes which are fundamental to galaxy formation.

\end{abstract}




\section{Introduction}

The Standard Cosmological model predicts that the vast majority of the matter is in the form of Dark Energy and Dark Matter. Only the remaining 4\% is in the form of baryons, the "normal matter" which makes stars and galaxies. The total baryon density of the Universe, $\Omega_{\rm baryons}$=0.0455, is well constrained from
measurements of Cosmic Microwave Background (CMB) anisotropies \citep{Aghanim16}, light element
abundances coupled with Big Bang nucleosynthesis \citep{Cooke18} and, in the near future, from observations of dispersion measures (DM) of Fast Radio Bursts (FRBs)  \citep[e.g.][]{Qiang20,Macquart20}. We note that results from these different experiments involving
vastly different physical processes and observation techniques agree remarkably well. However, 30 to 50\% of the baryonic matter is currently not observed directly \citep{Persic92, Nicastro18, Peroux20}. The latest baryon census shows that galaxies, groups and clusters together comprise only $\sim 10\%$ of the expected baryon density \citep{Salucci99,Bregman07,Shull12,Nicastro18, deGraaff18}. Recently, a
fraction of these cooler baryons were found on the scales of the galaxy haloes, including in the Circumgalactic Medium (CGM)
\citep{Werk13, Tumlinson17}. This still leaves a deficit of observed baryons relative to the predicted baryon density referred to as the "missing baryons problem" \citep{Bregman07, Shull12}.

At the turn of the century, cosmological hydrodynamical simulations of the large-scale structure have postulated on the phase and location of these missing baryons. At high-redshift, the models indicate that the majority of the baryons are in the low-density ionised gas of the intergalactic medium (IGM). This phase is observationally traced by the Ly-$\alpha$ forest which only provides direct constraints on the neutral fraction of a gas which is mostly ionised \citep{Kim13}. Therefore, baryons in IGM, although dominant at early epochs, remain challenging to observe. At lower redshifts, simulations of the matter distribution indicate that the baryons are related to a shock-heated phase of the gas with temperature range of $10^5$ < T < $10^7$ K, swiftly called Warm-Hot Intergalactic Medium (WHIM) \citep{Cen&Ostriker99, Dave01}. According to these cosmological simulations, at z = 0, the WHIM is the dominant baryon contributor and is located mostly in filaments and knots along the cosmic web, in which groups and clusters reside \citep{Martizzi19}. A different approach using \textit{Enzo} grid-code simulations challenges the postulate that a significant amount of WHIM has been shock-heated to 10$^7$ K. For example, \cite{Shull12} find a small "hot plume" at T$>$ 10$^7$ K in the (T,b) phase diagram. Observational baryon-censuses also find that the Ly-$\alpha$ forest contains $\sim$ 30\% of the baryons, with similar amounts in gas at T= 10$^{5-6}$ K probed by O\,VI and broad Ly-$\alpha$ absorbers (BLAs). Speculation is that the remainder (“missing baryons”) resides at higher temperatures (T $>$ 10$^{6}$ K) which are not traced by O\,VI and BLAs.


Due to its diffuse nature, 
direct detection of the WHIM in emission poses a great
observational challenge \citep{Bertone08, Frank10, Augustin19, Wijers20}. To remedy this limitation, absorption lines detected against bright
background objects offer the most compelling way to study the tempearture, ionisation state and column density of
the gas. In these absorbers, the minimum gas density detected is set by the brightness of the background
source and thus the detection efficiency is independent of redshift and the foreground object's brightness. 

Oxygen is one of the most useful tracers of the WHIM because of its high abundance and
because O\,VI, O\,VII, and O\,VIII excitation states span nearly the entire 10$^5$--10$^7$ K temperature range. The doublet of C\,IV and Ly-$\alpha$ contribute at the low end as well, while Ne\,VIII (at 770 and 780 \AA) is associated with collisionally ionised gas at T $\approx 5 \times10^5$ K \citep{Narayanan11}. The O\,VI doublet is observable in UV with
lines at 1032\AA\
and 1038\AA, which also makes them easy to identify \citep{Howk09}. Unfortunately, the interpretation of O\,VI is complex. One issue is that
O\,VI absorption turns out to be relatively easy to produce either from photoionization or collisional ionization (CIE).
In CIE, the maximum fraction of Oxygen which is in O\,VI is only $\sim$ 20\%, so the
ionization correction are likely considerable \citep{Werk14, Prochaska17}. Additionally, we need to specify a metallicity in
order to convert from 
O\,VI into a baryon mass, which is usually unknown. The WHIM can 
be traced by Ly-$\alpha$ absorption as well. These Ly-$\alpha$ lines look different from the narrow
lines typical of the photoionized gas, since they are also appreciably thermally broadened.
These are the BLAs, and overlap with the O\,VI-traced WHIM \citep[][]{Pachat16}.

Hotter collisionally ionized gas emits soft X-rays, but the emissivity is proportional to
the square of the gas density, which again limits the effectiveness of emission as a probe of low-density
gas. X-ray analysis traces metal emission lines (for the CGM, primarily O\,VII and O\,VIII, but also Iron lines).
Owing to the lack of telescopes with
adequate sensitivity in the X-ray wavelengths, the hot phase of the CGM traced by O\,VII and O\,VIII is as yet not well studied \citep{Li20}.  These lines, which correspond to a series of soft X-ray features around 0.6-0.8 keV, are commonly detected at redshift zero, corresponding
to the hot halo around the Milky Way \citep{Gupta12} (but see also \cite{Wang&Yao12}). At moderate
redshifts, they are also accessible with the grating spectrographs on {\it XMM-Newton} \citep{Arcodia18} and {\it Chandra} (sensitive down
about to 0.35 and 0.2 keV respectively). Unfortunately, current X-ray  gratings have poor spectral resolution (700-800km s$^{-1}$), and effective areas
of tens of cm$^2$ (compared to 1000-3000 cm$^2$ for {\it HST-COS} in the UV), so they have limited sensitivity
to the WHIM \citep{DeRoo20}. The detection of two X-ray absorbers towards 1ES 1553+113 by \cite{Nicastro18}
\citep[but see][for cautionary notes]{Johnson19} is the latest indication that the missing baryons are
indeed in a diffuse WHIM phase, although it remains unclear which fraction of the missing baryons
are traced by X-ray absorbers. Overall, there are only a few X-ray observed WHIM absorbers \citep{Bonamente16, Mathur17, Nicastro18, Johnson19, Nevalainen19}.
The next generation X-ray
observatories such as {\it Athena} \citep{Barcons11} and its high spectral resolution micro-calorimeter with about an order of magnitude higher effective area than {\it XMM-Newton} and {\it Chandra}, will
make significant progress in the next decade \citep{Barret20}. Until then, it is extremely important
to study the warm-hot gas, 
using the existing UV and X-ray
facilities.

Different methods have been proposed to detect the hot, highly-ionized WHIM gas: detection in galaxy groups with Sunyaev-Zeldovich effect
\citep{Hill16, Graaff19, Lim20,Tanimura20}
using autocorrelation function measurements \citep{Galeazzi10}, with absorption lines in quasar sightline \citep{Kovacs19} and using Cosmic Microwave Background (CMB) as a backlight \citep{Ho09}. 
Given the challenges of X-ray data, observations at longer wavelengths (UV and optical) benefit from higher instrumental throughpout, enhanced spectral resolution. By reverting to ground-based facilities, longer exposure times and larger number of targets become possible. Nevertheless, the UV lines have so far mostly been used to detect absorbing gas
with temperature range $\mathrm{10^5 < T < 10^6 K}$ from either O\,VI  \citep{Tripp00, Danforth&Shull05, Tripp08, Danforth&Shull08, Werk14, Savage14, Danforth16, Kacprzak16} or BLAs \citep{Lehner07, Danforth10}. Recently, \cite{Zastrocky18} have constrained the Milky Way's hot   
(T = 2 $\times$ 10$^6$ K) coronal gas using the forbidden 5302\,\AA\ transition of Fe\,XIV. 
Only recently, some highly ionised iron UV lines detected in emission have been used as diagnostics of gas at temperatures of T=10$^7$ K. Out of several forbidden lines in the UV that could trace this gas temperature range, and from various species of highly ionised iron, the emission of [Fe\,XXI] is the brightest \citep{Anderson16}. 
\cite{Anderson18} report the discovery of [Fe\,XXI] in emission in a filament projected 1.9 kpc from
the nucleus of M87. 
   Theoretically, the highly ionised iron UV lines can be observed in absorption as well. The forbidden line of [Fe\,XXI], in particular, has the largest effective cross-section
for absorption and a rest wavelength $\lambda$1354\,\AA, conveniently close to Ly-$\alpha$ $\lambda$1215\,\AA.





In this work, we target the [Fe\,XXI] line in absorption as a tracer of high-redshift 10$^7$K WHIM gas. Large optical spectroscopic quasar surveys available nowadays offer a new opportunity to statistically probe these filaments by detecting the matter between the high-redshift background quasar and the observer. Here, we make use of a large sample of high-spectral resolution quasar spectra with known intervening neutral gas Damped Ly-$\alpha$ Absorbers (DLAs). Recent studies \citep{Peroux19, Hamanowicz20} provide evidence for the paradigm of the origin of DLAs, showing that these objects probe overdensities (e.g. groups) in the Universe. We use these systems as tracers of foreground oversdensities, and to increase the sensitivity of the experiment we stack multiple quasar spectra at the DLA's redshifts \citep{Ribas17} to look for indications of T=10$^7$K WHIM gas traced by the [Fe\,XXI] absorption line. 


This manuscript is organized as follows: Section 2 presents the observations used for this study. Section 3 details the stacking method, while Section 4 summarizes our findings and future prospects with upcoming facilities. Throughout
this paper,  we adopt an H$_0$ =70 km s$^{-1}$ Mpc$^{-1}$, $\Omega_M$ =0.3, and $\Omega_{\Lambda}$ = 0.7 cosmology.


\section{A large sample of high resolution quasar spectra}

\begin{figure}
\centering
\includegraphics[width=1.1\columnwidth]{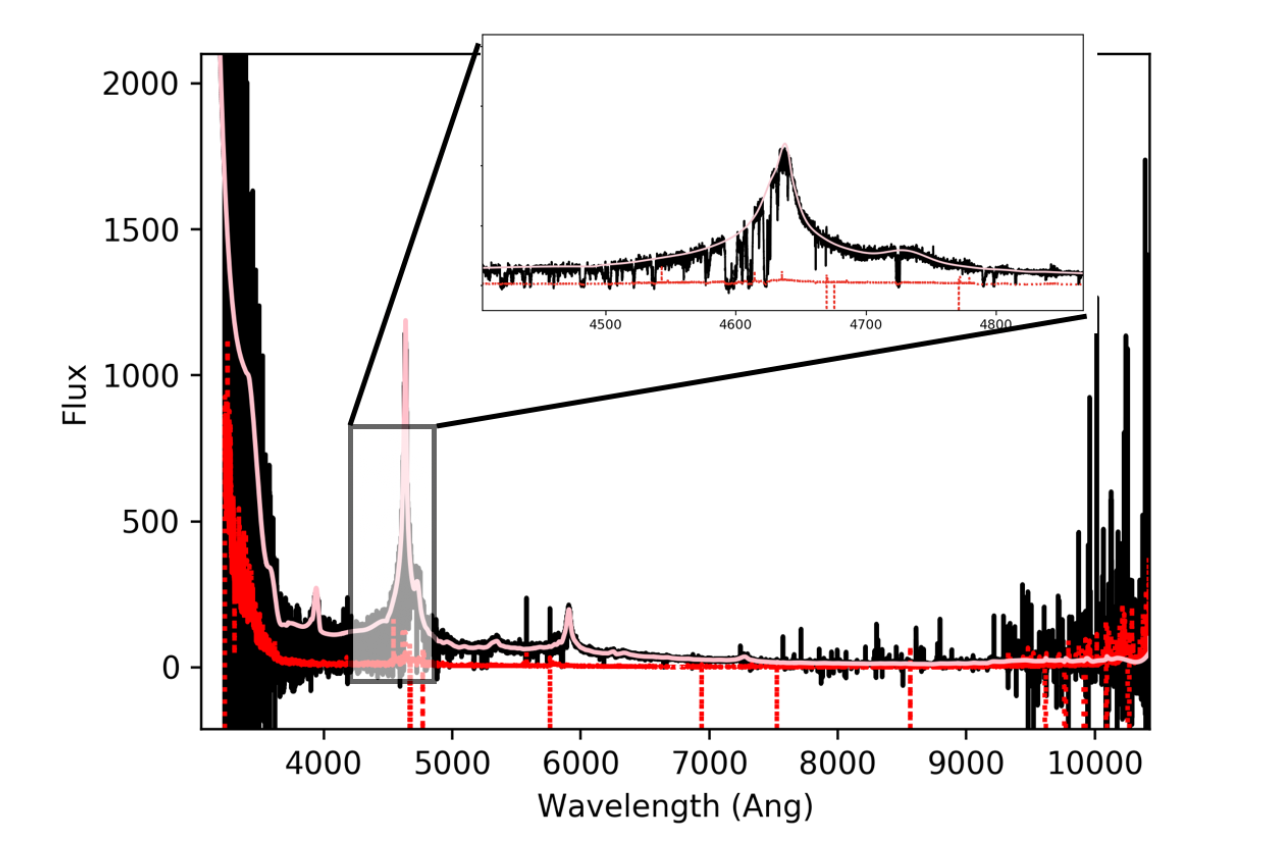}
\caption{Example VLT/UVES quasar spectrum. The black histogram shows the full quasar spectrum at observed wavelengths and with arbitrary flux units.
The quasar J000149-015939 at z$_{\rm em}$=2.815 has a strong Ly$\alpha$ emission line at $\sim$ 4650\,\AA\ at restframe. The red histogram displays the corresponding error array. The pink line indicates the fitted quasar continuum. The inset zooms around the Lyman-$\alpha$ emission revealing the high $snr_{ind}$.}
\label{fig:full_spec}
\end{figure}

To enhance the sensitivity to absorption line detection, this study focuses on the quasar spectra with the highest spectral resolution available. 
This work is based on a quasar sample observed with the Ultraviolet and Visual Echelle Spectrograph (UVES) of the European Southern Observatory's (ESO's) 8-metre Very Large Telescope (VLT) \citep{Dekker00}. The UVES instrument is a two-arm (red and blue) grating cross-dispersed echelle spectrograph. 
Some observations are made using only one arm, but most observations are done using both arms simultaneously. The light is then split in two by a dichroic mirror and centered around to a standard or user-defined central wavelength.

\begin{figure}
\centering
\includegraphics[width=\columnwidth]{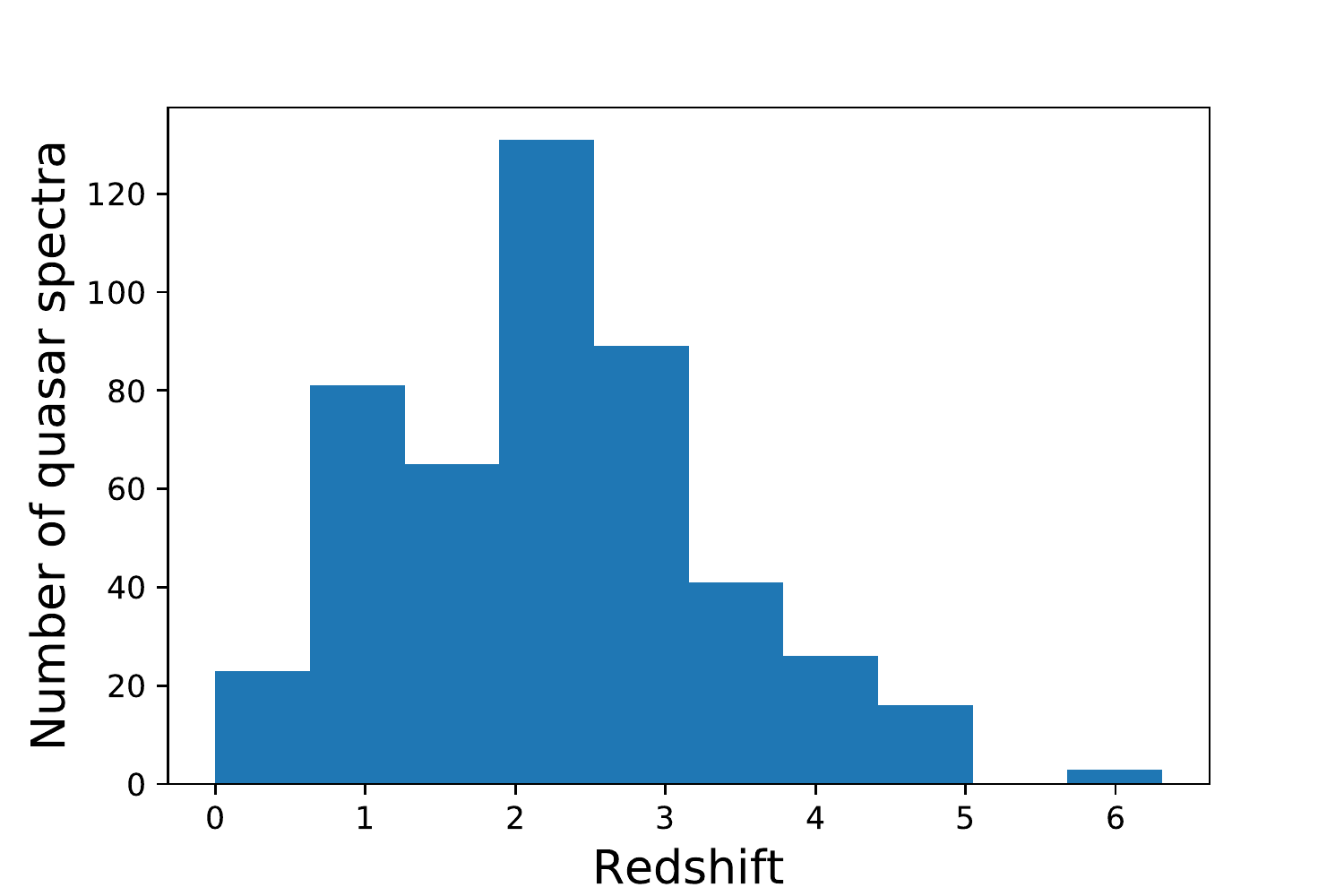}
\caption{Quasar emission redshift distribution. The 467 UVES quasar spectra from the SQUAD sample cover a broad redshift range from $\mathrm{z_{em}= 0-6}$, with wavelength coverage from 3,050\AA to 10,500\AA\ with gaps depending on the chosen spectral settings. }
\label{fig:histo_zQ}
\end{figure}

\begin{figure}
\centering
\includegraphics[width=\columnwidth]{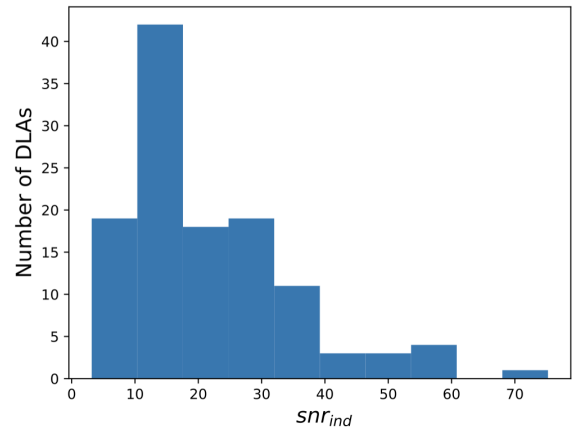}
\caption{Signal-to-noise per resolution element ($snr_{ind}$) distribution. The histogram shows the 155 UVES quasar spectra containing at least one DLA. The mean resolving power of the UVES spectrograph (R > 40,000) combined with an averaged exposure time of 5.9 hours result in sizeable $snr_{ind}$, reaching up to $snr_{ind}$>70.}
\label{fig:histo_SNR}
\end{figure}

Our study makes use of the first release of the "Spectral Quasar Absorption Database" (SQUAD) with 467 fully reduced, continuum-fitted quasar spectra \citep{Murphy19}.
Figure~\ref{fig:full_spec} shows an example of a VLT/UVES quasar spectrum, with a zoom on the Lyman-$\alpha$ quasar emission line. In this sample, the spectra were cross-matched with observations from the ESO UVES archives to include all quasars from literature up to August 2017 \citep{Flesch15}. The redshift was also cross-matched with three databases (SDSS, NED and SIMBAD). In cases where no matches were found, a measure of the redshift was derived directly from the UVES spectrum. Figure~\ref{fig:histo_zQ} illustrates the quasar redshift distribution of the whole sample. The redshifts range from $\mathrm{z_{em}= 0-6}$, with a broad wavelength coverage range (3,050\AA-10,500\AA) with gaps depending on the chosen spectral settings. The mean resolving power of the exposures is
R = 49,000. The corresponding resolution element is 6.1 km s$^{-1}$. The quasar spectra have on average a total exposure time of 5.9 hours, resulting on sizeable signal-to-noise ratios  per resolution element (hereafter $snr_{ind}$, for individual quasar spectra), reaching up to $snr_{ind}$>70. In other words, the full quasar sample is made of over 1450hrs of VLT time observations. Figure~\ref{fig:histo_SNR} displays the  distribution of the $snr_{ind}$ of the UVES quasar spectra.

\begin{figure}
\centering
\includegraphics[width=\columnwidth]{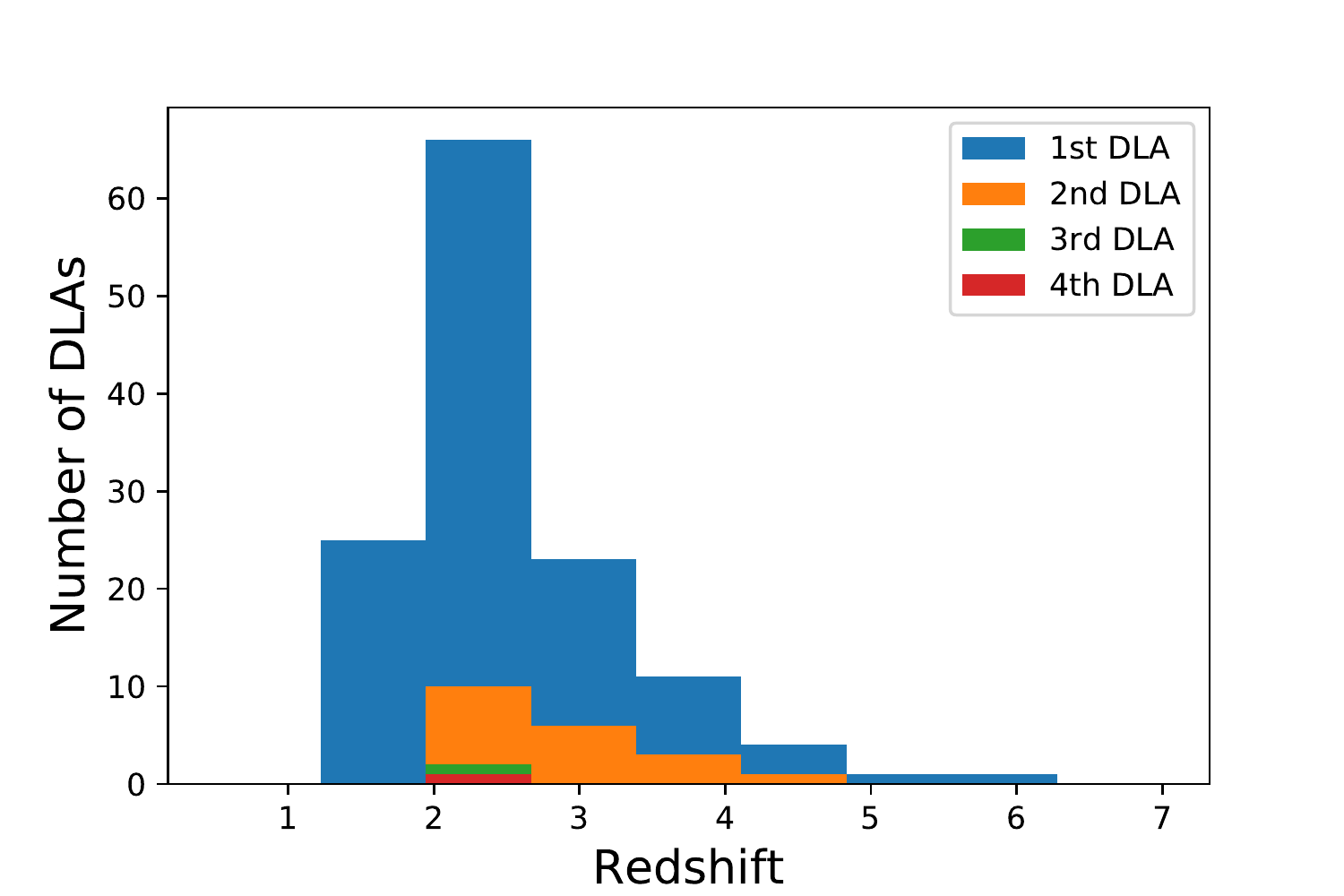}
\caption{Redshift distribution of the 155 DLAs. Orange (green/red) histograms indicate the redshift of 2$\mathrm{^{nd} (3^{rd}/4^{th}}$) DLA in one given quasar spectrum. The mean DLA redshift for the sample is ${z_{DLA}= 2.5}$.}
\label{fig:histo_zDLA}
\end{figure}

Additionally, \cite{Murphy19} provide a catalogue of 155 identified Damped Lyman-$\alpha$ Absorbers (DLA), out of which 18 are reported for the first time. The newly identified DLAs were found by visually checking the UVES quasar spectra. These DLAs are a class of quasar absorbers tracing high H\,I column density cold (T=$\mathrm{10^4K}$) gas, with N(H\,I)$\geq 2*10^{20}$ \citep{Wolfe86}. These systems comprise the majority of the neutral gas reservoir in the Universe used for the initial phase of star formation (e.g \cite{Peroux20}). Additionally, their metal content provides crucial information about the chemical evolution of galaxies. Figure~\ref{fig:histo_zDLA} displays the redshift distribution of this sample of 155 DLAs. Orange (green/red) histograms indicate the redshift of $\mathrm{2^{nd} (3^{rd}/4^{th})}$ DLA in those quasar spectra containing more than one. The mean DLA redshift for the sample is ${z_{DLA}= 2.5}$.

\section{A stacked high-resolution DLA spectrum}
\label{sec:maths} 

At the high-resolution (R > 40,000) offered by UVES, the metallicity of individual DLAs is routinely well-measured \citep[e.g.][]{Kulkarni05, Decia18, Poudel20}. In this work, we target weak metal lines which are not expected to be detected in a single spectra. In particular [Fe\,XXI] has an oscillator strength $f_{osc} = 5.3 \times 10^{-6}$ and a Doppler parameter $b_{Fe} = (2kT/m_{Fe})^{1/2} \approx (54.4 km s^{−1}) (T_{7})^{1/2}$ where $T_{7}$ indicates at T = 10$^{7} $K. In order to probe these significantly weaker absorption lines, we build a stack of more than hundred high-resolution quasar spectra, shifted at the DLA redshifts.

\subsection{Selecting quasar spectra}

First, targeting DLAs to search for metal lines limits the number of quasar spectra to be stacked to the ones containing known DLAs. This step results in a sample of 155 quasar spectra out of the 467 contained in the whole SQUAD quasar sample. 

Second, the dispersion of the UVES spectra, expressed in km/s per pixel, differs between five different values ranging from 1.3 km/s per pixel to 2.5 km/s per pixel. In order to prevent rebinning of the spectra during the stacking procedure, we elect to only include spectra with identical dispersion. A total of 137 spectra 
have a dispersion of 2.5 km/s per pixel which are then used in the subsequent study.

Finally, the spectral gaps in between the settings are evidenced by a visual inspection of the spectra. This gaps appear as flat lines in the middle of different parts of the spectra, where there is no signal. Only spectra with the appropriate wavelength coverage were included in further calculations depending on the location of the spectral gaps with respect to the targeted element at a given absorption redshift. This again caused a reduction of the number of spectra stacked depending on the element under study and its corresponding observed wavelength, where the count of the number of UVES spectra included for each element is stated on every plot.
For each metal line stacked, we compute the number of spectra covering the appropriate wavelength given the DLA's redshift and the quasar spectrum wavelength coverage (in the rest-frame ranging typically from $\mathrm{\lambda1,200}$ to $2,600$\,\AA). The number of spectra included in the stack of each metal line are therefore different. For example, [Fe\,XXI] has a final number of stacked spectra of 106.

Note that some quasars contain multiple DLAs at different redshifts (with cases containing up to four DLAs in one quasar spectrum), so some quasar spectra are used multiple times. 


\subsection{Stacking Procedure}

We first shift the selected spectra to the DLAs rest-frame wavelength. We recall that all spectra have the same dispersion of 2.5 km/s per pixel, so that no rebinning is required in building the stack in velocity space. We then use the resulting quasar spectra sample to compute a median stack. We reject spectra which do not fully cover the wavelength range $\pm$500 km/s from the targeted metal feature. The corresponding number of quasar UVES spectra utilised are given in the legend of each of Figures~\ref{fig:FeII}, \ref{fig:FeXXI} and \ref{fig:StrongLinesI}. At each pixel, we calculate the middle value of the ordered array of normalised fluxes from each quasar spectra. We also perform mean and weighted-mean stacks. Our results indicate that the median stack recovers the normalised quasar continuum better, with a higher resulting $SNR_{stack}$ (for signal-to-noise of stack spectra) and less prominent contamination from interlopers. The median stack therefore provides the most stringent limits and is used in the subsequent analysis.

\subsection{Strong Metal Lines Detections}

The neutral gas probed by DLAs contains a wide variety of metals with different ionization states. These are tracing gas temperature of the order 10$^4$K \citep{Tumlinson17}.
The metallicity of the DLAs evolves from $Z = 0.003 Z_{\odot}$ solar at redshift $z =4$, to $Z = 0.15 Z_{\odot}$  solar at redshift $z=1$ \citep{Rafelski12, Decia18, Poudel20, Peroux20}, with no DLAs found with a metallicity below 0.0025 solar \citep{Meiksin09, Cooke18}.

\begin{figure}
\centering
\includegraphics[width=\columnwidth]{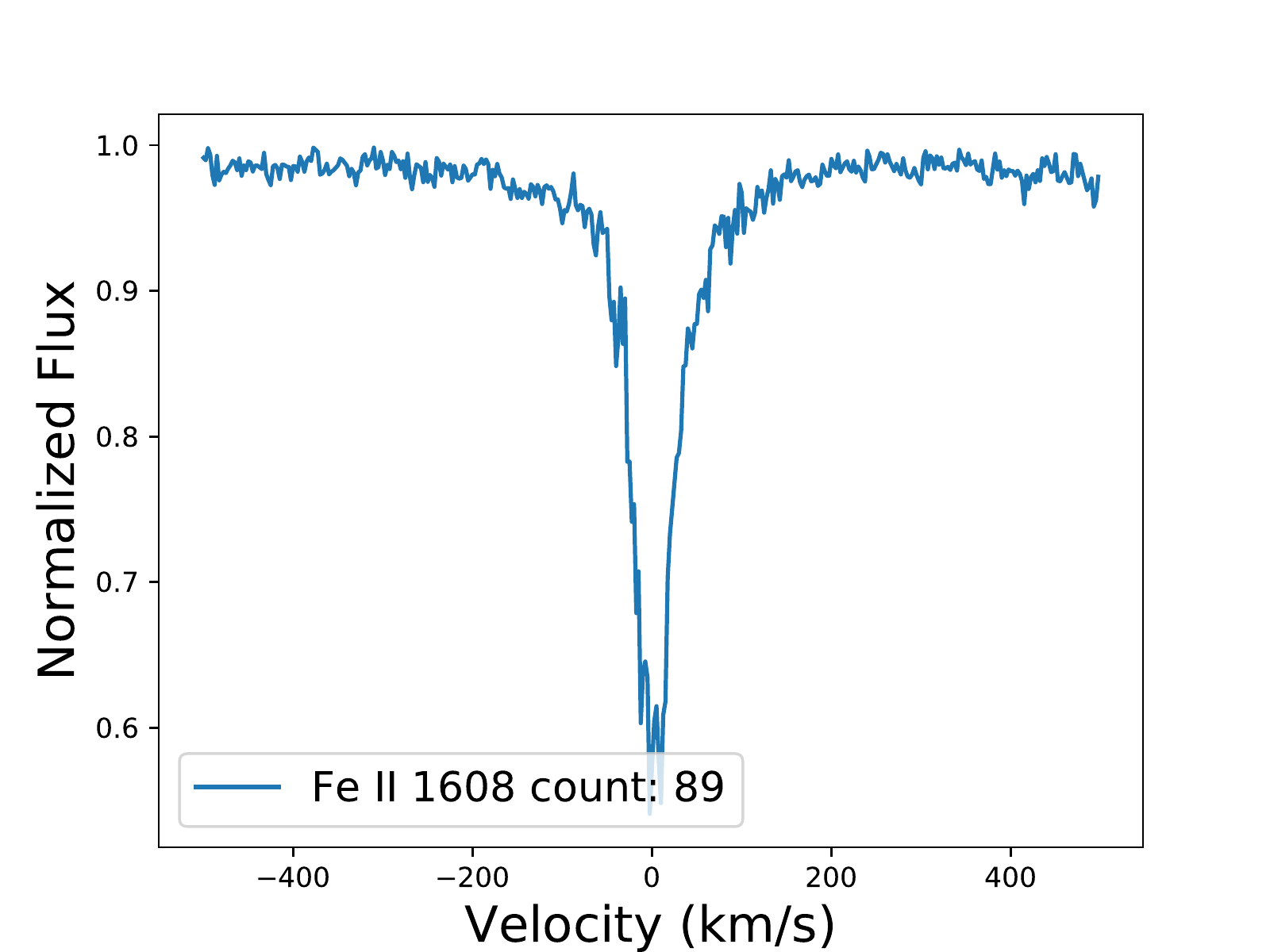}
\caption{Example stack of a strong iron line. Stacking of 89 (as indicated in the legend) normalised UVES quasar spectra at the DLA position displays the detection of the Fe II $\lambda_{\rm rest}$1608\ line.}
\label{fig:FeII}
\end{figure}

The median stack spectrum results in the detection of several strong metal absorption lines commonly reported in DLAs which demonstrates the sensitivity of our approach. Figure~\ref{fig:FeII} shows an example of the median stack of 89 DLA spectra. The strong iron absorption line, Fe\,II ($\lambda_{\rm rest}$=1608), is undoubtedly detected. 
The following 15 lines were detected at the redshift of the DLAs with a wavelength interval between 1,200\,\AA\ to 2,600\,\AA: iron (Fe II 1608\,\AA, Fe II 2600\,\AA), silicon (Si  II 1260\,\AA, Si IV 1393\,\AA, Si IV 1403\,\AA), carbon (C IV 1548\,\AA, C IV 1551\,\AA), sulphur (S II 1259\,\AA), zinc (Zn II 2026\,\AA), oxygen (O I 1302\,\AA), chromium (Cr II 2026\,\AA), manganese (Mn II 2577\,\AA), nickel (Ni II 1317\,\AA), aluminium (Al II 1671\,\AA, Al III 1863\,\AA). These detections are reported in Figure~\ref{fig:StrongLinesI} of Appendix~\ref{sec:appendix}.

\section{Probing the Warm-Hot Intergalactic Medium}


\subsection{Observed [Fe\,XXI] Column Density Limit}
\label{sec:col_limit}

\begin{figure}
\centering
\includegraphics[width=\columnwidth]{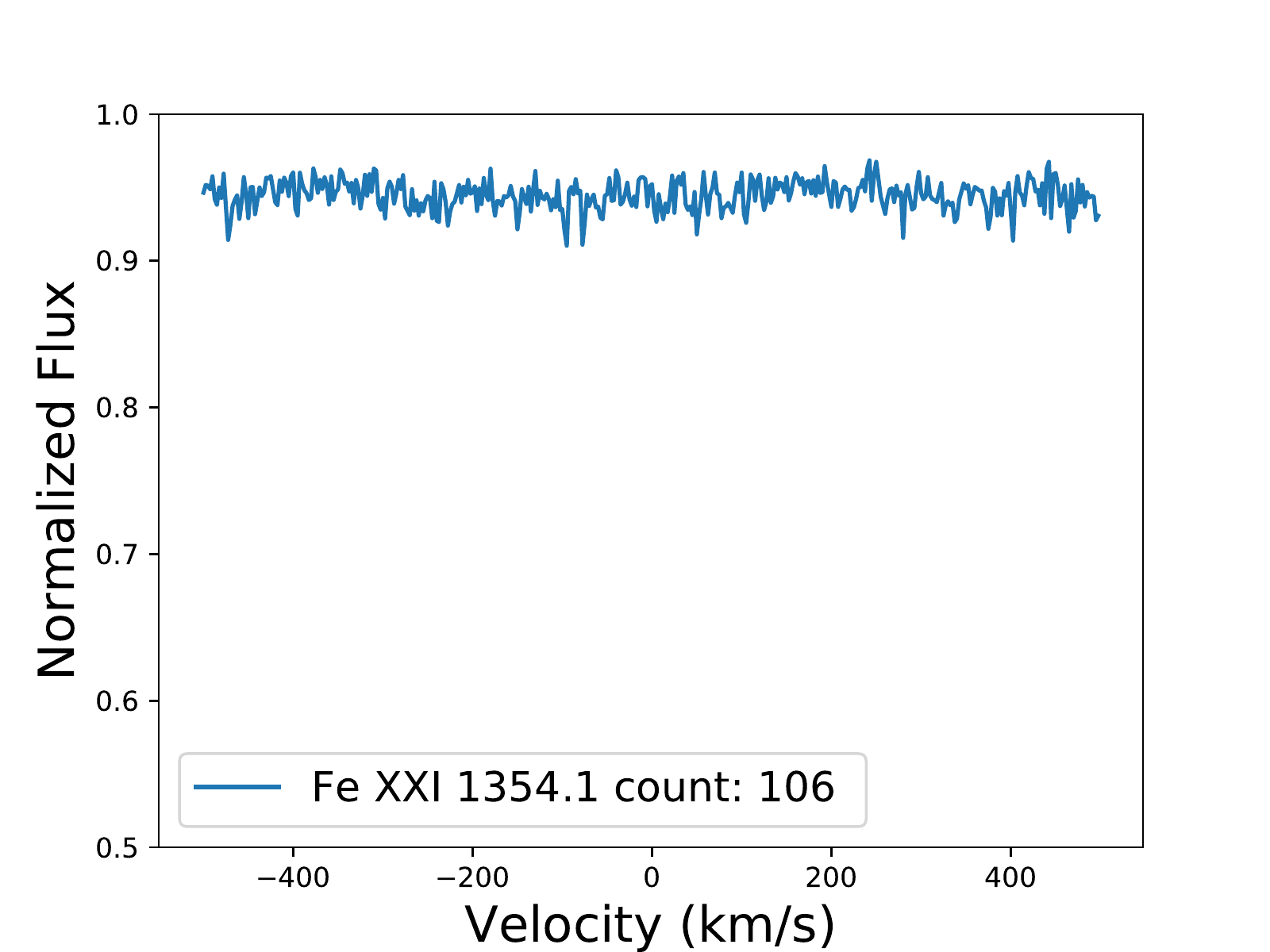}
\caption{Stack of the weak [Fe\,XXI] line. Stacking of 106 (as indicated in the legend) normalised UVES quasar spectra at the DLA position lead to a non-detection of Fe\,XXI 1354\AA. The y-axis values are set to match those of Figure~\ref{fig:FeII}}.
\label{fig:FeXXI}
\end{figure}

To search for tracers of the WHIM at UV-wavelengths, we also stack forbidden UV lines of highly ionized iron which trace T=10$^7$ K gas. The strongest of these line is [Fe\,XXI] at rest wavelength $\lambda$1354.1 \citep{Anderson16}. This line is not detected in the stack spectrum as illustrated in Figure~\ref{fig:FeXXI}. Another 15 weak metal lines were searched for but not detected. These include absorption lines like chlorine, argon, titanium, chromium, cobalt, germanium, arsenic, and krypton \citep{Prochaska03, Ellison10}.

To compute the equivalent width (EW) upper limit for this non-detection, we use the following relation \citep{Menard03}:

\begin{equation}
\mathrm{EW_{obs}< \bigg[\frac{\sigma \times FWHM}{ SNR_{stack}}\bigg] \approx (77 m \si{\angstrom} )\bigg(\frac{\sigma}{5}\bigg)\bigg(\frac{FWHM}{1.43}\bigg)^{-1} \bigg(\frac{SNR_{stack}}{93}\bigg)^{-1}}
\label{eqn:Ew}
\end{equation}

where we assume a $\sigma$ value equal to 5, and compute the signal-to-noise ratio ($SNR_{stack}$=93) of the stacked spectrum in the velocity range ${-500 < v < 500 }$ km/s. The expected [Fe\,XXI] line is broad at the observed wavelength 4739 \AA\ (1354.1\AA\ at a mean redshift of <z>=2.5). At T = 10$^7$ K,
the Doppler parameter b = 54.4 km/s and the FWHM = 2(ln 2)$^{1/2}$b = 90.6 km/s for a Gaussian line profile \citep{Danforth10, Keeney12}. We therefore assume a FWHM = 1.43 \AA\ in equation~\ref{eqn:Ew}. We compute the minimum observed equivalent width (${EW_{obs}<77}$m\AA) and convert it to rest-frame equivalent width ${EW_{rest}<22} $m\AA. We compute the column density according to the linear relation between the equivalent width and the column density:

\begin{equation}
    \mathrm{EW_{rest}}= \bigg(\frac{\pi e^{2}}{m_{e}c}\bigg) \frac{Nf_{osc}\lambda_{rest}^{2}}{c} \approx (0.860 \mathrm{m} \si{\angstrom})\bigg[\frac{N({\mathrm{[Fe\,XXI]}})}{10^{16}\mathrm{cm}^{-2}}\bigg]
    \label{eqn:ew}
\end{equation}

where c is the speed of light, $m_e$ the electron mass,
$\lambda_{rest}$ is the rest wavelength of [Fe\,XXI], $\mathrm{\lambda_{rest}=1354}$\,\AA\ and $\mathrm{f_{osc}=5.3\times 10^{-6}}$ is the oscillator strength for [Fe\,XXI]. 

Reverting the equation, we compute the corresponding column density $\mathrm{log[N([Fe\,XXI])]}$ using the value of $\mathrm{EW_{rest}}$ ( ${EW_{rest}<22}$m\AA) as follows:

\begin{equation}
    N({[Fe\,XXI]}) = (10^{16} cm^{−2})(EW_{rest}/0.860 m\si{\angstrom})
    \label{eqn:col}
\end{equation}

The resulting column density limit in the stack of 106 DLAs observed in UVES quasar spectra at a mean redshift of ${z_{DLA}= 2.5}$ is $\mathrm{log[N([Fe\,XXI])]<17.4}$, where all column densities are expressed in units of $cm^{-2}$.


\subsection{Expected [Fe\,XXI] Column Density}

In order to put these results in context, we make two estimates of the expected column density of [Fe\,XXI] in gas typical of the WHIM and its corresponding \textit{EW}. In both cases, we assume that [Fe\,XXI] corresponds to 20\% of the total iron abundance in the gas, i.e. an ionisation fraction  of (Fe$^{\rm +20}$/Fe)=0.2. This hypothesis is born from modelling of gas in collisional ionization equilibrium (CIE) using the CHIANTI database \citep{Anderson16}.Then we first assume the metallicity of the gas to be Z$_{\rm Fe}$=0.1 Z$_{\odot}$, i.e. $\mathrm{(Z_{\rm Fe}/Z_\odot)= 10^{-1}}$.  Given a specific element, $Fe$, we refer to the usual relation \citep[e.g.][]{Peroux20}:

\begin{equation}
 \mathrm{N_{FeXXI} = N_{HI} \bigg(\frac{Fe}{H}\bigg)_{\odot} \bigg(\frac{Fe^{+20}}{Fe} \bigg) \bigg(\frac{Z_{Fe}}{Z_\odot} \bigg) = 3.55 \times 10^{14} cm^{-2}}
\label{eqn:metallicity}
\end{equation}

where N(Fe) indicates the column density of the element Fe (surface density in atoms $\mathrm{cm^{-2}}$), $\mathrm{(Z_{\rm Fe}/Z_\odot)}$ is the logarithmic abundance relative to the assumed solar abundance, and $\mathrm{{N(Fe)}/{N(H)}_{\odot}}$ is the reference solar abundance of both the element and hydrogen in the Solar System. Here, we adopt values of the Solar System abundances of \cite{Asplund09} resulting in a column density of iron $\mathrm{log[{N(Fe)}]_{\odot} = 7.5}$ and column density of hydrogen $\mathrm{log [{N(H)}]_{\odot} = 12}$, meaning that log(Fe/H)$_\odot$=$-4.50$.  

Using the mean hydrogen column density of the DLAs in the sample: $\mathrm{log[N(H\,I)]=20.75}$, we calculate by assuming $\mathrm{(Z_{\rm Fe}/Z_\odot)= 10^{-1}}$ that the WHIM gas will have an expected column density of $\mathrm{log[N([Fe\,XXI])] \sim 14.5}$. In a second hypothesis, we assume that the metallicity of the WHIM gas is $\mathrm{(Z_{\rm Fe}/Z_\odot)= 10^{-2}}$, which then leads to an expected column density of $\mathrm{log[N([Fe\,XXI])] \sim 13.5}$. These column densities are about three orders of magnitude smaller than the upper limit derived from the stack of 106 UVES quasar spectra: $\mathrm{log[N([Fe\,XXI])]<17.4}$. 

We note that in collisional ionisation equilibrium \citep{Shull82}, the ion fraction $f_{[Fe\,XXI]} = (Fe^{+20}/Fe)$ $\sim$ 0.246 peaks at logT = 7.0, but falls off rapidly at lower temperatures: $f_{[Fe\,XXI]} = 0.148$ (log$T$ = 6.90) and $f_{[Fe\,XXI]} = 0.026$ (logT = 6.80). Similar fall-off occurs at higher temperatures. Thus, the observed limits on $N([Fe\,XXI])$ are for a narrow temperature range, logT$ = 7.0 \pm 0.1$. The failure to detect [Fe\, XXI] could mean that the WHIM associated with the DLAs is cooler than T=$10^{6.8}$K. We note however that several recent studies have shown that DLA systems are primarily found in foreground overdensities (e.g. groups), which likely contains logT = 7.0 gas \citep{Peroux19, Hamanowicz20}. Alternatively, to be able to detect such [Fe\,XXI] column density, one would need to significantly increase the number of spectra in order to get a higher SNR$_{stack}$.

\subsection{Future Prospects of Probing the WHIM at UV wavelengths}

We now assess if this experiment with upcoming or planned facilities would enable to reach the [Fe\,XXI] column density to trace expected from the WHIM. First, we use equation~\ref{eqn:ew} to compute the rest equivalent width corresponding to $\mathrm{log[N([Fe\,XXI])] \sim 14.5}$. We derive $\mathrm{EW_{rest}= 0.03}$m\AA. Assuming a mean DLA redshift for the sample $\mathrm{z_{DLA}= 2.5}$, we compute a corresponding observed equivalent width of typically $\mathrm{EW_{obs}= 0.1}$m\AA. 

We look in turn into three type of facilities with high-resolution spectroscopic capabilities. First, we focus on the Multi-Object Spectrograph Telescope (4MOST), a high-multiplex spectroscopic survey instrument currently under construction phase for the 4m VISTA telescope of the European Southern Observatory (ESO). 4MOST is expected to start its science operations in 2023. It has a wide field of view of 2.5 degrees diameter and nearly 2,400 fibers dividing in two different types of spectrographs. We here focus on the high-resolution fibers, offering nearly 800 spectra with a spectral resolution R$\sim$20,000 (FWHM=0.2\AA\ at 6000\AA) over a wavelength coverage $\lambda 3920$ to $6750$\ \citep{Quirrenbach15}. For a typical exposure time of 2hrs, 4MOST will deliver quasar spectra with an average $snr_{ind}$=10 for a mag $<$19 object \citep{deJong19}. We note that the WEAVE instrument designed for the William Herschel Telescope (WHT) will offer comparable characteristics in the northern hemisphere \citep{Pieri16}. 

Secondly, we consider the High Resolution spectrograph (HIRES) for the 39m Extremely Large Telescope (ELT) currently under construction. HIRES \citep{Marconi18}, which has successfully completed its Phase A, has a large spectral coverage ranging from 4000 \AA\ to 25000 \AA\ with a spectral resolution R=100,000 (FWHM=0.04\AA\ at 6000\AA). HIRES is not a survey instrument, as opposed to the other facilities described here,  being both a single-target spectrograph and a facility open to the community for small and medium-size proposals. However, the ESO open-archives policy \citep{Romaniello18} means that significant number of quasar spectra will become publicly available after few years of operations. Depending on the science goals, we expect quasar spectra to be recorded with $snr_{ind}$ varying from 10 to 1000. Here, we assume a medium $snr_{ind}$ per spectra of $snr_{ind}$=50.

Finally, we look further out in the future for upcoming facilities for multi-object spectrographs on 10m class telescopes. The Maunakea Spectroscopy Explorer (MSE) is being proposed to replace the Canada-France-Hawaii Telescope (CFHT) \citep{MSE19}. With 1.52 deg$^2$ field of view, it will have the capability of simultaneously observe more than 4,000 objects. At low spectral resolution ($R \sim 3,650$), a $snr_{ind}$ of 2 for magnitude 24 sources will be achieved in one hour observation. At high spectral resolution ($R \sim 40,000$; \textit{FWHM}=0.1\,\AA\ at 6000\,\AA), a $snr_{ind}$=20 for magnitude 20 source will be obtained typically in a little over 1-hour exposure. A conceptual design study for an analogous facility in the southern hemisphere has also been put forward. The so-called Spectroscopic Survey Telescope, here after SpecTel \citep{Ellis17}, will offer a field of view of 5 deg$^2$ and will be equipped with 15,000 fibers covering a wavelength range of $3,600 <\lambda<13,300$\,\AA.

\begin{table}
\begin{center}
\caption{Number spectra needed for each facility to reach the $SNR_{stack}$ necessary to detect [Fe\,XXI] in the WHIM at $\sigma=3$. 
The first two rows indicate the spectral resolution of the facilities expressed as \textit{R} and in \AA\ as \textit{FWHM} at 6000\,\AA. The third row provides the required \textit{SNR$_{stack}$} of the stack spectra to reach the ${EW_{obs}= 0.1}$m\,\AA\ required to detect [Fe\,XXI]. The fourth row states the mean $snr_{ind}$ of individual spectra foreseen for these facilities given a typical exposure time. Finally, the last row provides the number of quasar spectra required to achieve the corresponding \textit{SNR$_{stack}$} in the stack spectra. Note that while the unmatched high spectral resolution of ELT/HIRES leads to a smaller number of spectra, this facility does not provide the multiplexing capabilities of other telescopes (4MOST, MSE/SpecTel). 
}
\begin{tabular}{ccccc} 
\hline
\bf Facilities & \bf 4MOST & \bf ELT/HIRES  & \bf MSE / SpecTel  \\[0.2cm]
 \hline
  Spectral Resolution, R & 20,000 & 100,000 &  40,000\\[0.3cm]
 \hline
  FWHM [\AA] & 0.2 & 0.04 & 0.1 \\[0.3cm]
 \hline
  Required $SNR_{stack}$ to& 42,900 & 42,900 &  42,900 \\
 reach $\mathrm{Ew_{obs}= 0.1}$m\AA&  &  &   \\
 \hline
 Individual $snr_{ind}$ & 10 & 50 & 20 \\[0.3cm]
 \hline
 Number of Spectra  & $>$18M & $>$700k & $>$4M
\label{Numspec}
\end{tabular}
\end{center}
\end{table}

The characteristics of each of these facilities are summarised in the first three lines of Table~\ref{Numspec}. We next estimate the \textit{SNR$_{stack}$} of the stack quasar spectrum required to reach the expected observed equivalent width of  $\mathrm{EW_{obs}= 0.1}$m\AA\ to detect a $\mathrm{log[N([Fe\,XXI])] \sim 14.5}$ absorption feature at $\sigma=3$. To this end, we revert equation~\ref{eqn:Ew} and derive \textit{SNR$_{stack}$}=42,900. We note that given the width of the [Fe\,XXI] line (FWHM=1.43 \AA), this calculation is independent of the spectral resolution of the instrument. We then compute the number of spectra, \textbf{N}, needed for each facility to reach the necessary \textit{SNR$_{stack}$} in the stack spectrum that will allow one to detect [Fe\,XXI] in WHIM gas:


\begin{equation}
\label{solar}
N > \bigg[\frac{SNR_{stack}}{snr_{ind}}\bigg]^2
\end{equation}

We find that 4MOST, ELT/HIRES and MSE/SpecTel will require respectively $>$18M, $>$700k and $>$4M quasar spectra to achieve the required sensitivity. The high-multiplexing capabilities of 4MOST, MSE and SpecTel mean that large samples can be acquired in a few years, although these large number of spectra are unrealistic at present. The experiment we explore here would benefit from cumulating data from existing and upcoming quasar surveys \citep[e.g.][]{Liske08, Merloni19}.

To put these results in context, we stress that this experiment is complementary to approaches planned at X-ray wavelengths with up-coming facilities. In coming decades, next-generation spacecrafts will aim at solving fully the missing baryons problem. The ESA {\it Athena} mission with spectroscopic and imaging capabilities in the 0.5--12keV range will significantly further our understanding of the baryons in the Universe, both inside the potential wells of groups and clusters of galaxies and in the WHIM in filaments between the densest regions in the Cosmos \citep{Barret20}. Investigating how such potential wells formed and evolved, and how and when the material trapped in them was energised and chemically enriched, can uniquely be tackled by observations in the X-ray band, combining wide-field images with high resolution spectroscopy, both of high sensitivity. Similarly, {\it Lynx}, a mission proposed to NASA in the framework of the next flagship space telescope, will have total effective area greater than 2 m$^2$ at 1 keV \citep{Falcone18}. These facilities will provide a new window in our capability to characterise the physical processes of the WHIM. However, these exposures will reveal single WHIM detections, while the detection of [Fe\,XXI] in stack quasar spectra will offer a complementary way to characterise the physical properties of the WHIM in a statistical manner. By reverting to a statistical approach, this technique is less prone to uncertainties related to variation from object to object in the Universe (the so-called cosmic variance).

\section{Conclusion}

In this work, we target the [Fe\,XXI] line in absorption as a tracer of high-redshift 10$^7$K WHIM gas. We make use of a large sample of high-spectral resolution quasar spectra with known intervening neutral gas Damped Ly-$\alpha$ Absorbers (DLAs). To enhance the sensitivity to absorption line detection, this study focuses on high spectral resolution (R$\sim$49,000) quasar spectra from VLT/UVES. To increase the sensitivity of the experiment, we stack 106 quasar spectra at DLA's redshifts to look for indications of 10$^7$K WHIM gas traced by the [Fe\,XXI] absorption line. We report a limit on the column density in the stack of $\mathrm{log[N([Fe\,XXI])]<17.4 cm^{-2}}$. Using basic assumptions on properties of 10$^7$ K gas, we calculate that the WHIM will result in an expected column density of the order $\mathrm{log[N([Fe\,XXI])] \sim 14.5 cm^{-2}}$, almost three orders of magnitude smaller than the derived upper limit. We then analyse the capabilities of future facilities, namely 4MOST, ELT/HIRES and MSE/SpecTel to detect such gas. We compute the number of spectra that would be needed from each facilities to reach the  signal to noise ratio required for the detection of [Fe\,XXI] absorption line in 10$^7$K WHIM gas. We find that 18M high-resolution fibre 4MOST, 700k ELT/HIRES or 4M MSE/TechSpec quasar spectra are required. We have explored here an alternative method that this experiment could be initiated by cumulating samples from existing and planned quasar surveys with these facilities. In conclusion, we study here an alternative to X-ray detected WHIM tracers. While a significantly weaker line, redshifted [Fe\,XXI] 1354\,\AA\ is observable at optical wavelengths from the ground, benefiting from higher instrumental throughput, enhanced spectral resolution, longer exposure times and increased number of targets.


\section*{Acknowledgements}

We thank Michael Murphy and collaborators for making reduced and combined UVES quasar spectra publicly available to the community. We are grateful to the referee, Mike Shull, for a thorough report.

\section*{Data availability}
 The data underlying this article are available in [UVES\_SQUAD\_DR1], at  [doi:10.1093/mnras/sty2834]. The datasets were derived from sources in the public domain: [https://github.com/MTMurphy77/UVES\_SQUAD\_DR1;].




\bibliographystyle{mnras}
\bibliography{biblio}



\appendix

\section{Strong Metal Lines Detections}
\label{sec:appendix}

The figure in this appendix reports the detection of multiple strong metal lines in UVES stacked quasar spectra at the DLA redshift. 

\begin{figure*}
   \includegraphics[width=\columnwidth]{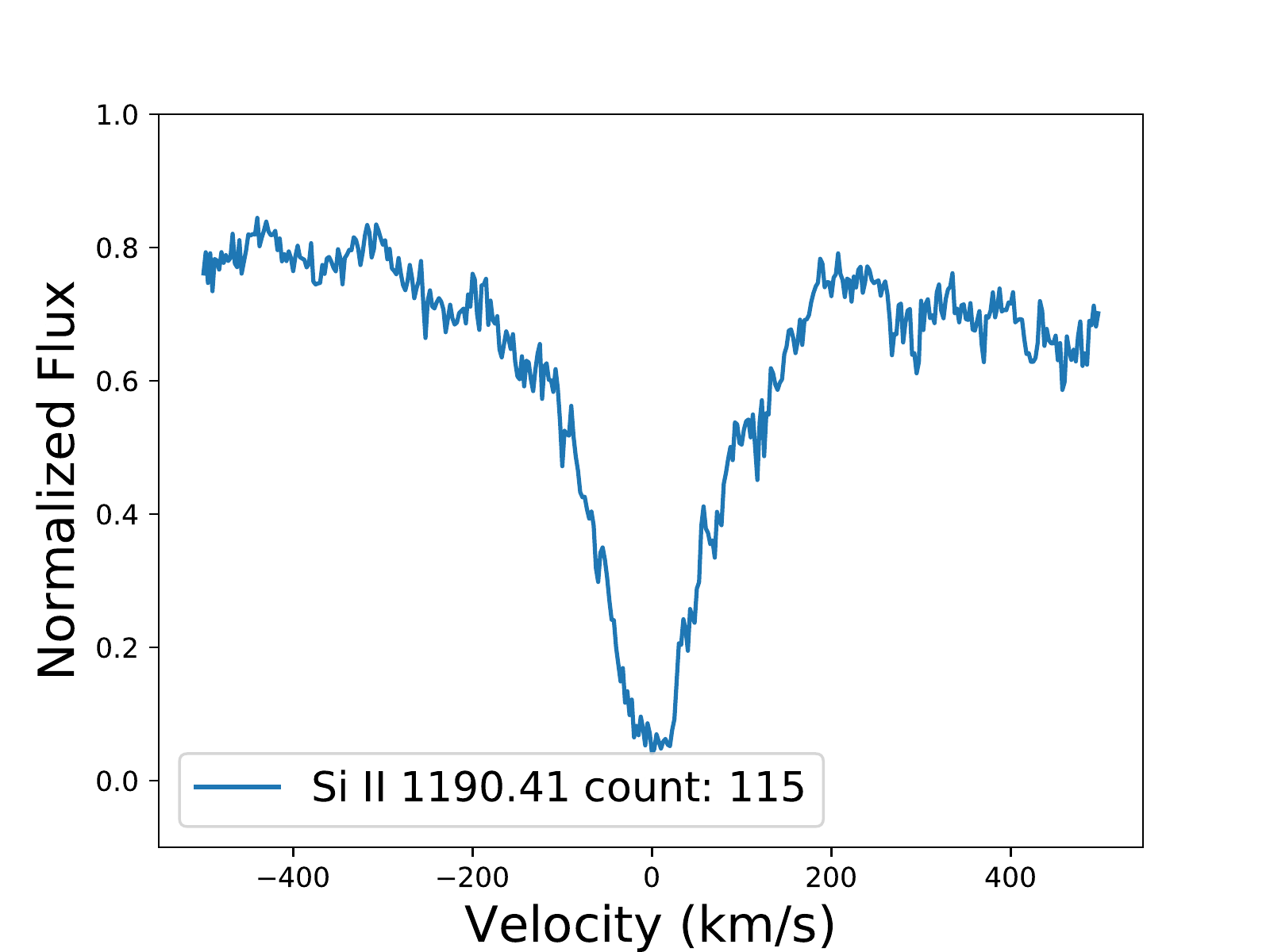} 
  \includegraphics[width=\columnwidth]{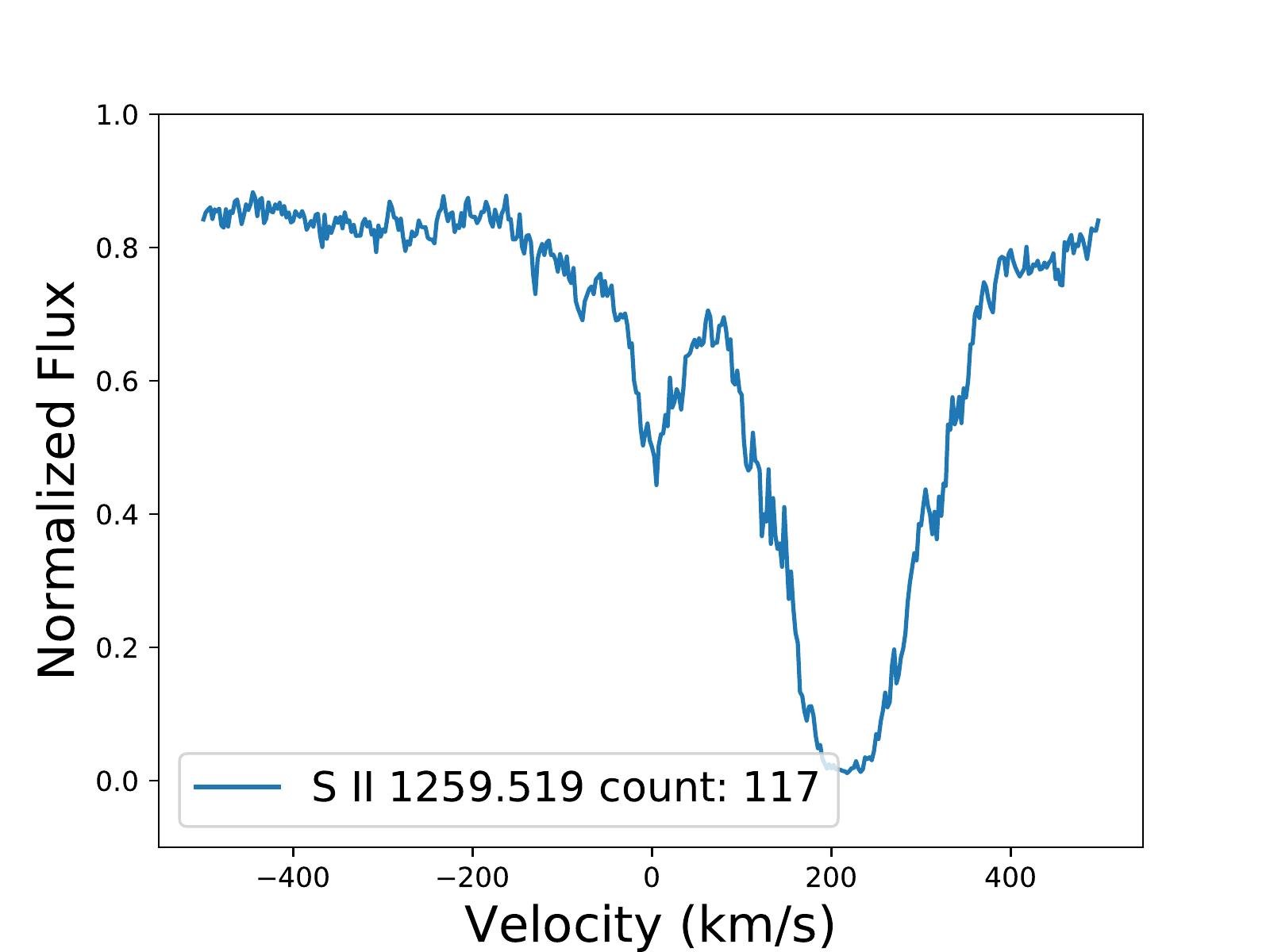}   
  \includegraphics[width=\columnwidth]{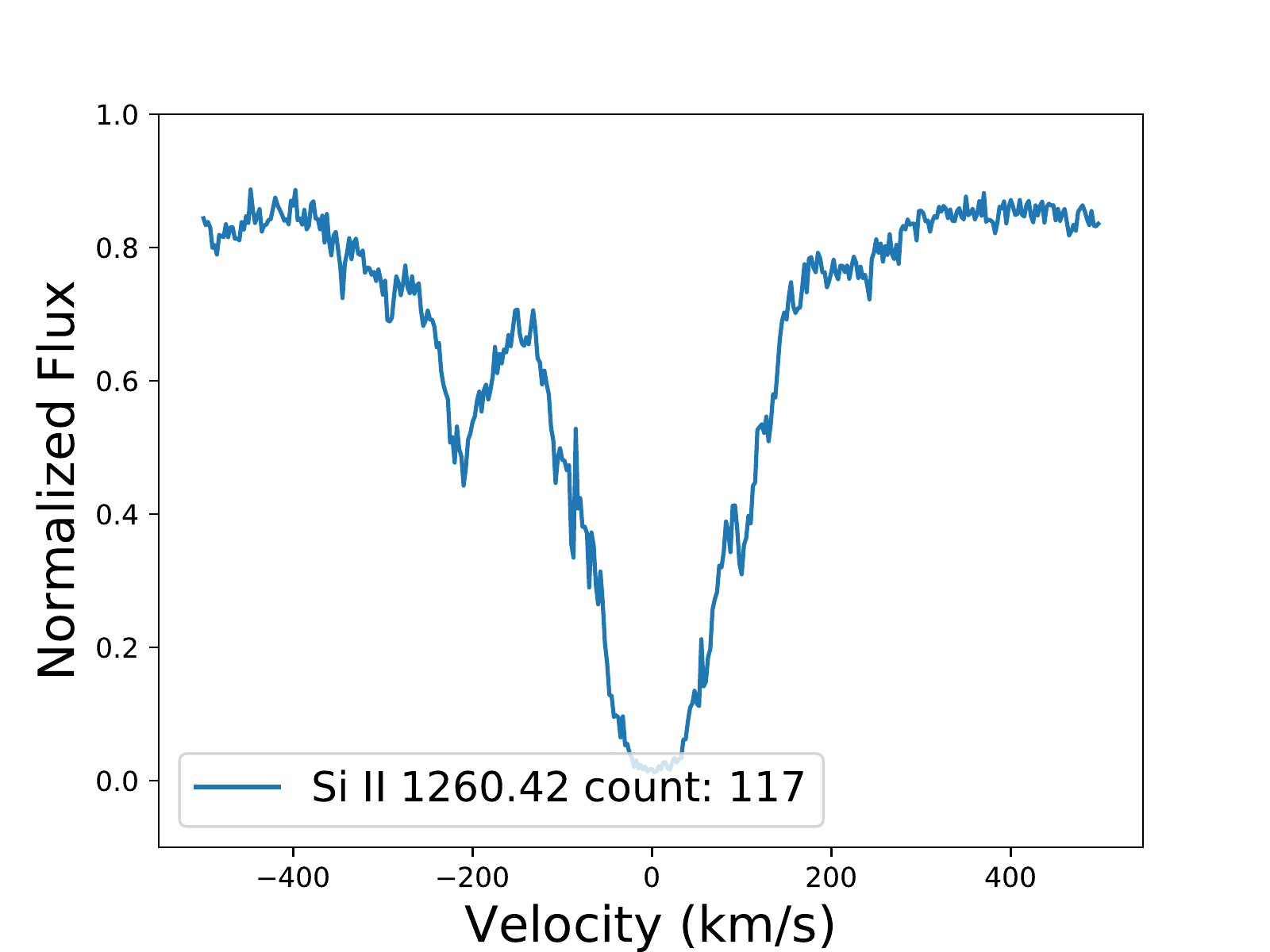} 
    \includegraphics[width=\columnwidth]{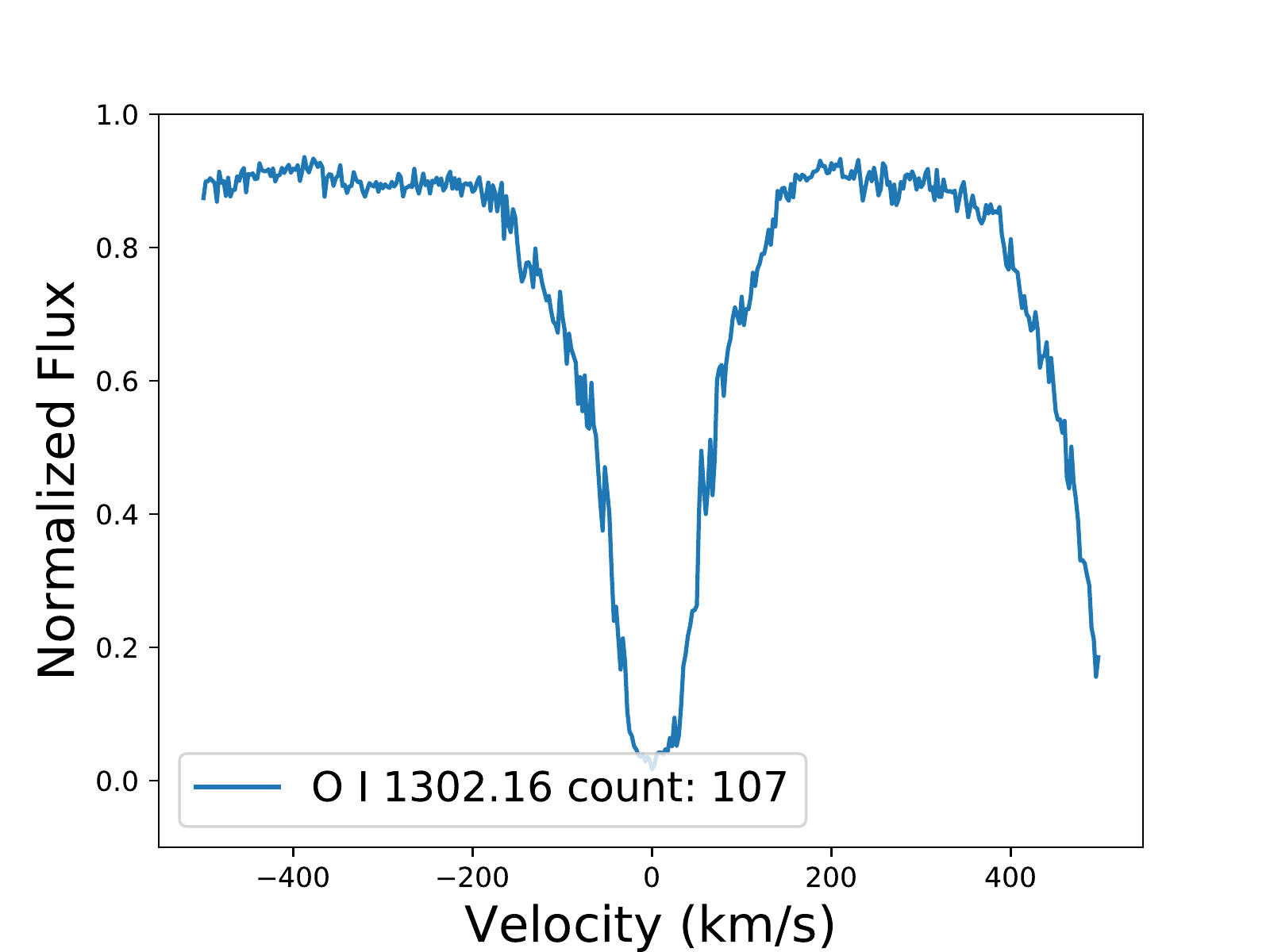} 
  \includegraphics[width=\columnwidth]{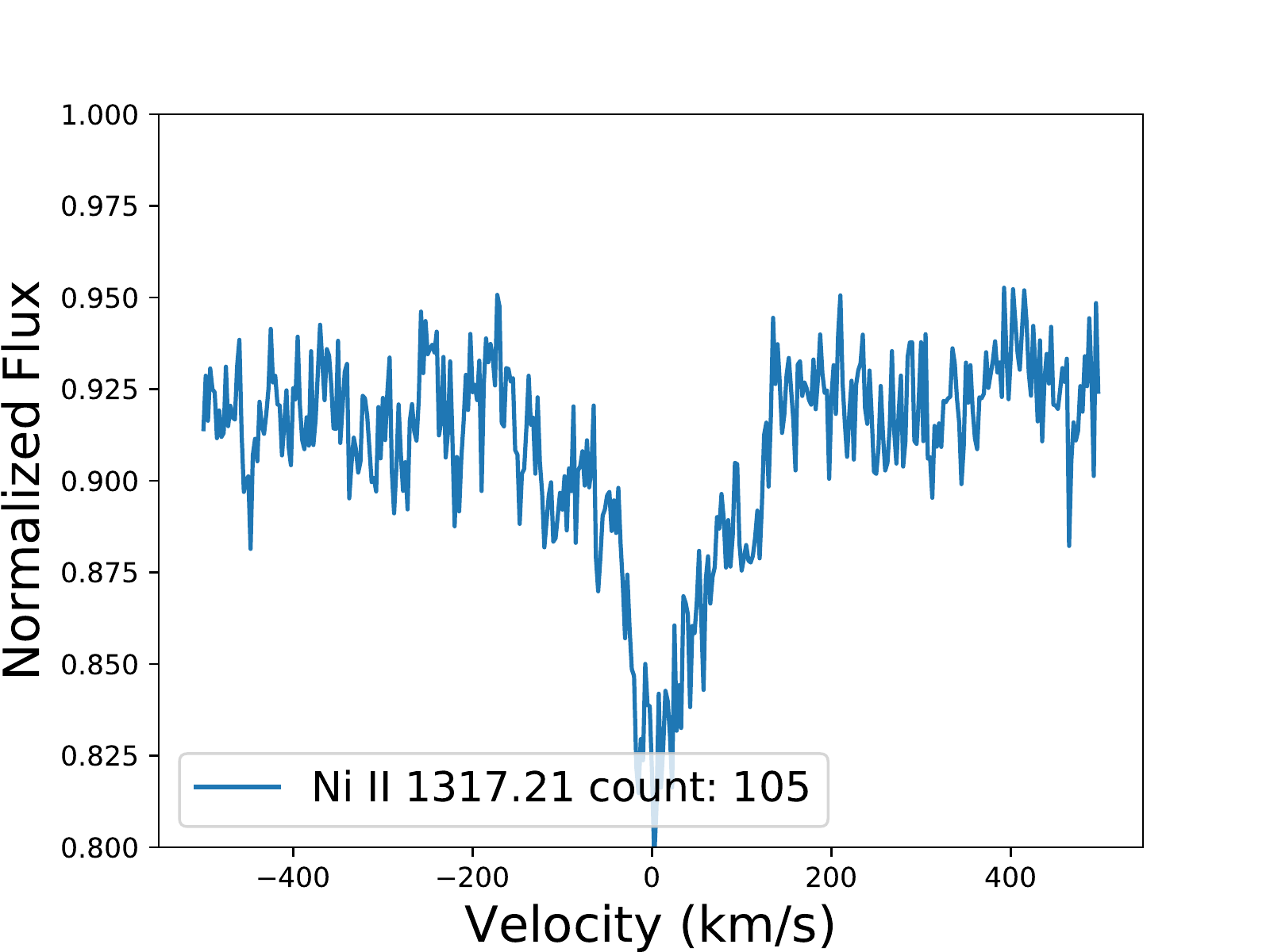} 
  \includegraphics[width=\columnwidth]{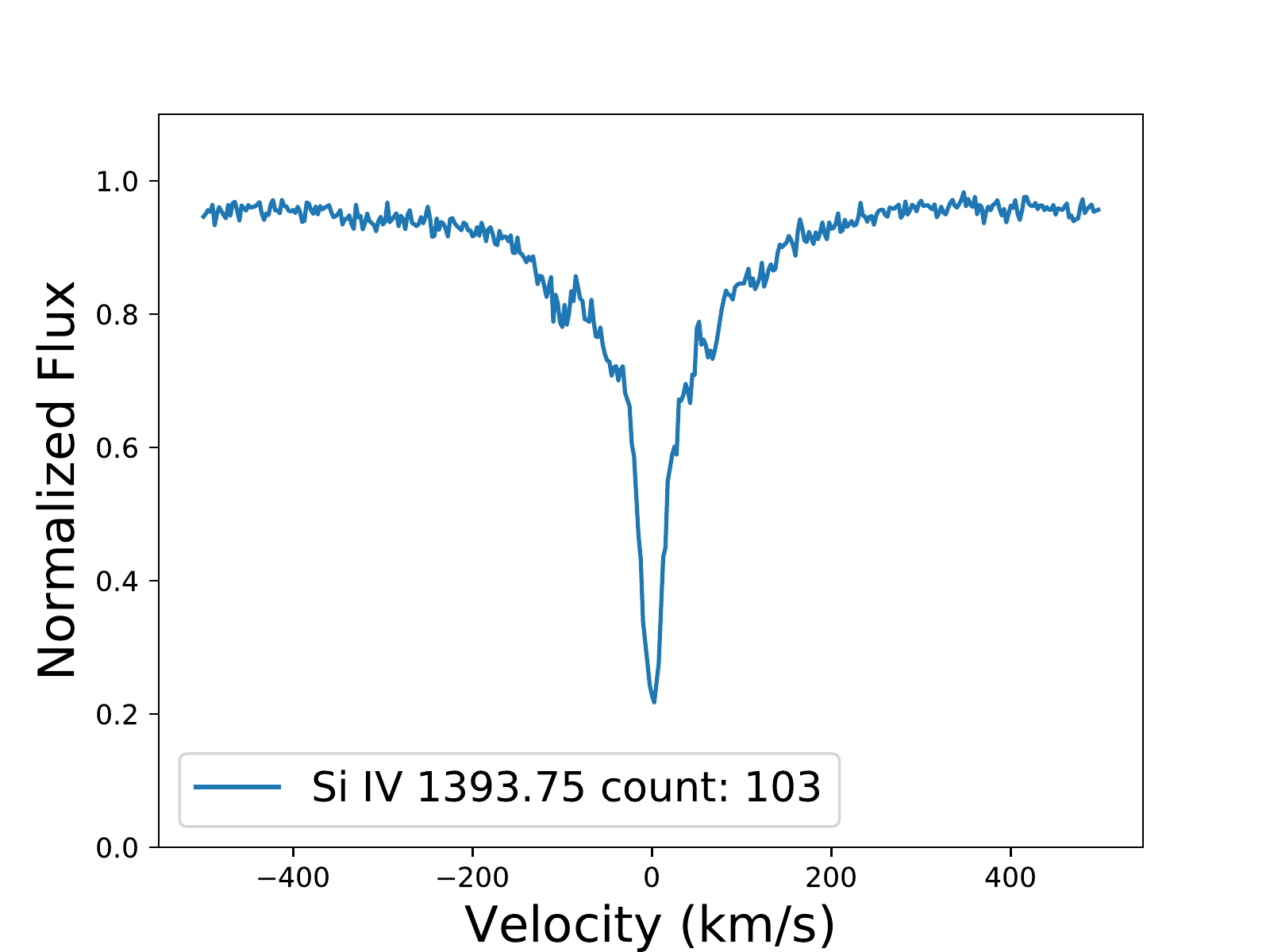}
  \includegraphics[width=\columnwidth]{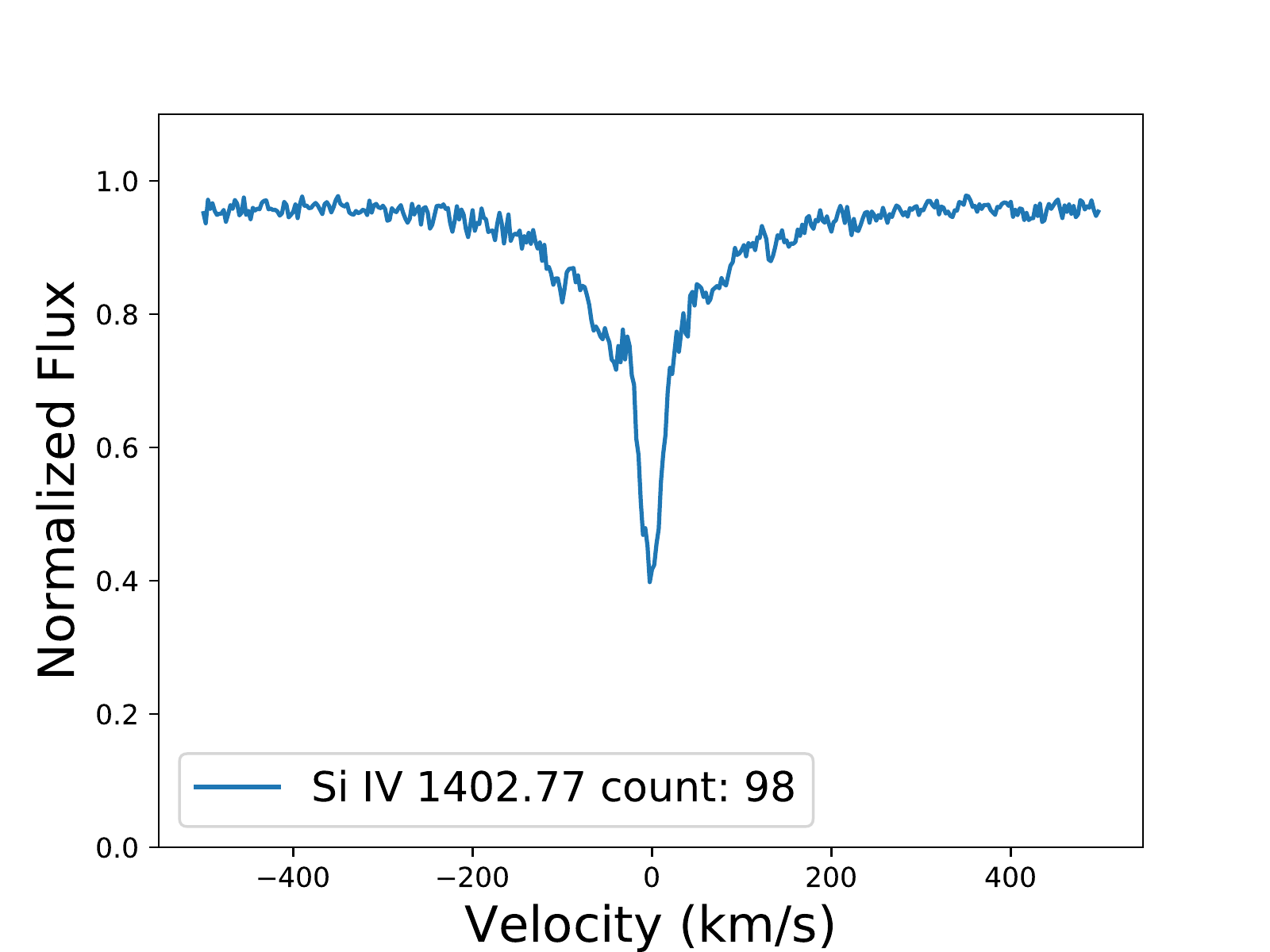} 
  \includegraphics[width=\columnwidth]{newfigs/Fe160.pdf}
\\
\end{figure*}%

\begin{figure*}
   \includegraphics[width=\columnwidth]{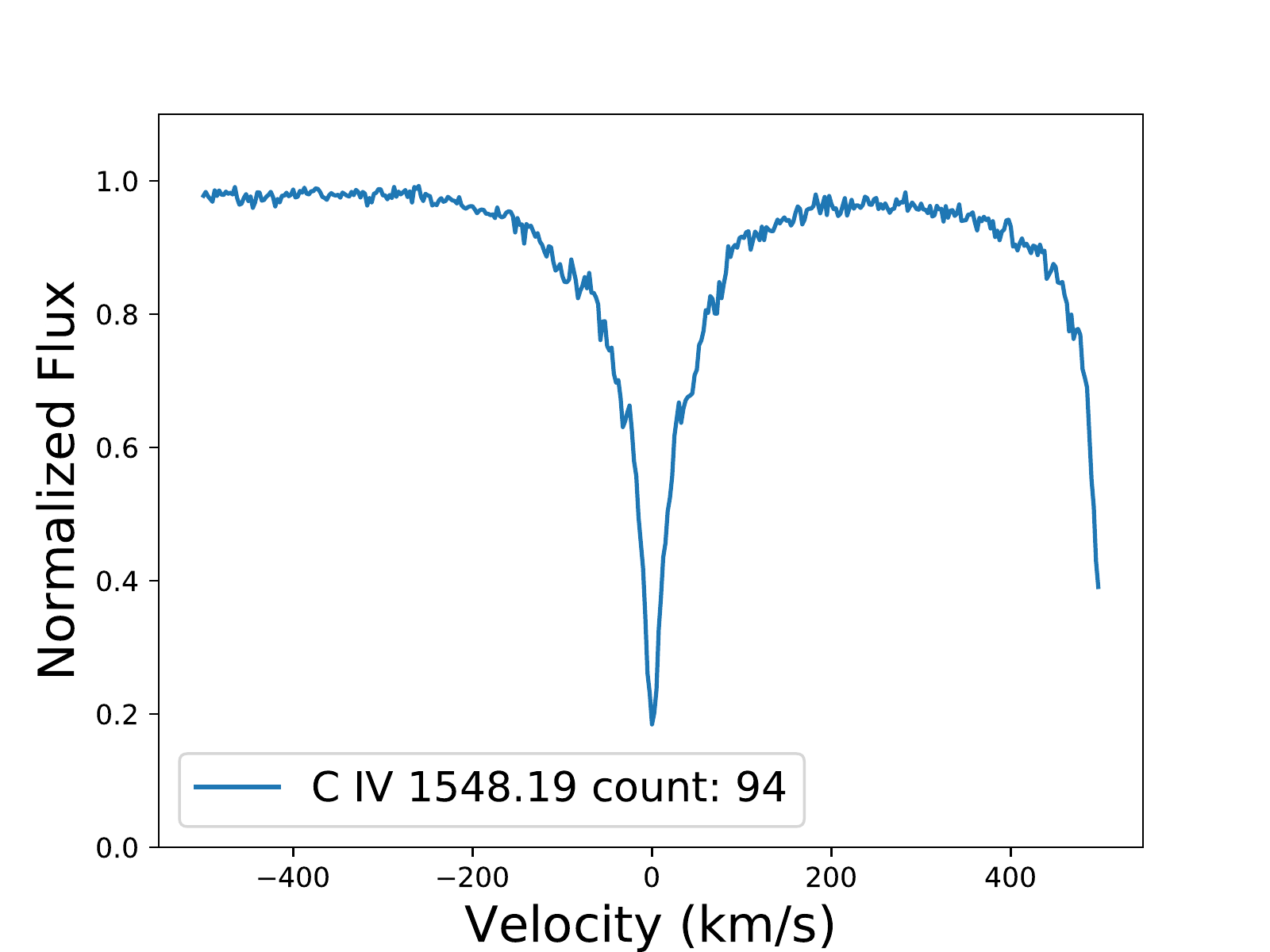} 
 \includegraphics[width=\columnwidth]{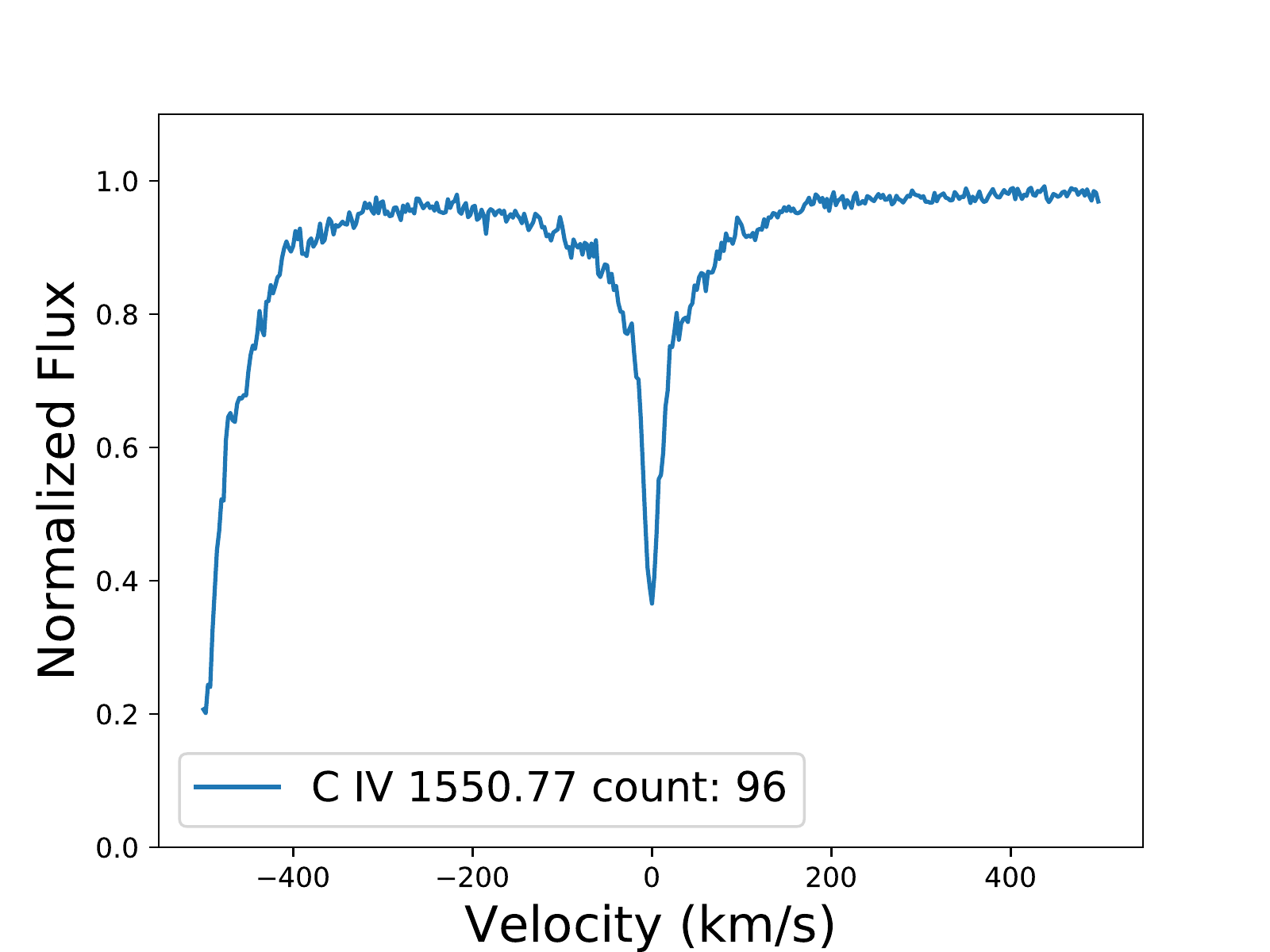}  
  \includegraphics[width=\columnwidth]{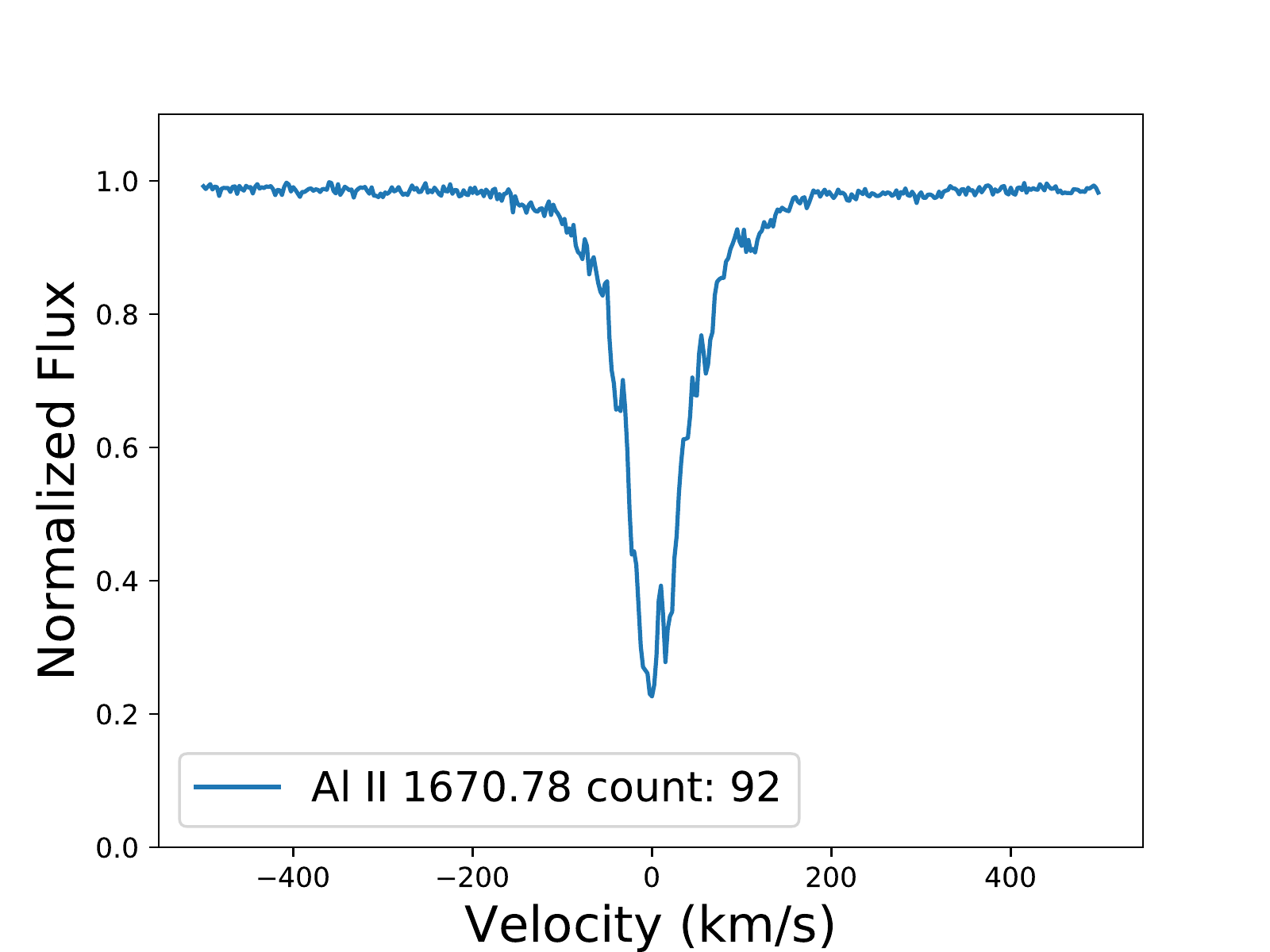}
   \includegraphics[width=\columnwidth]{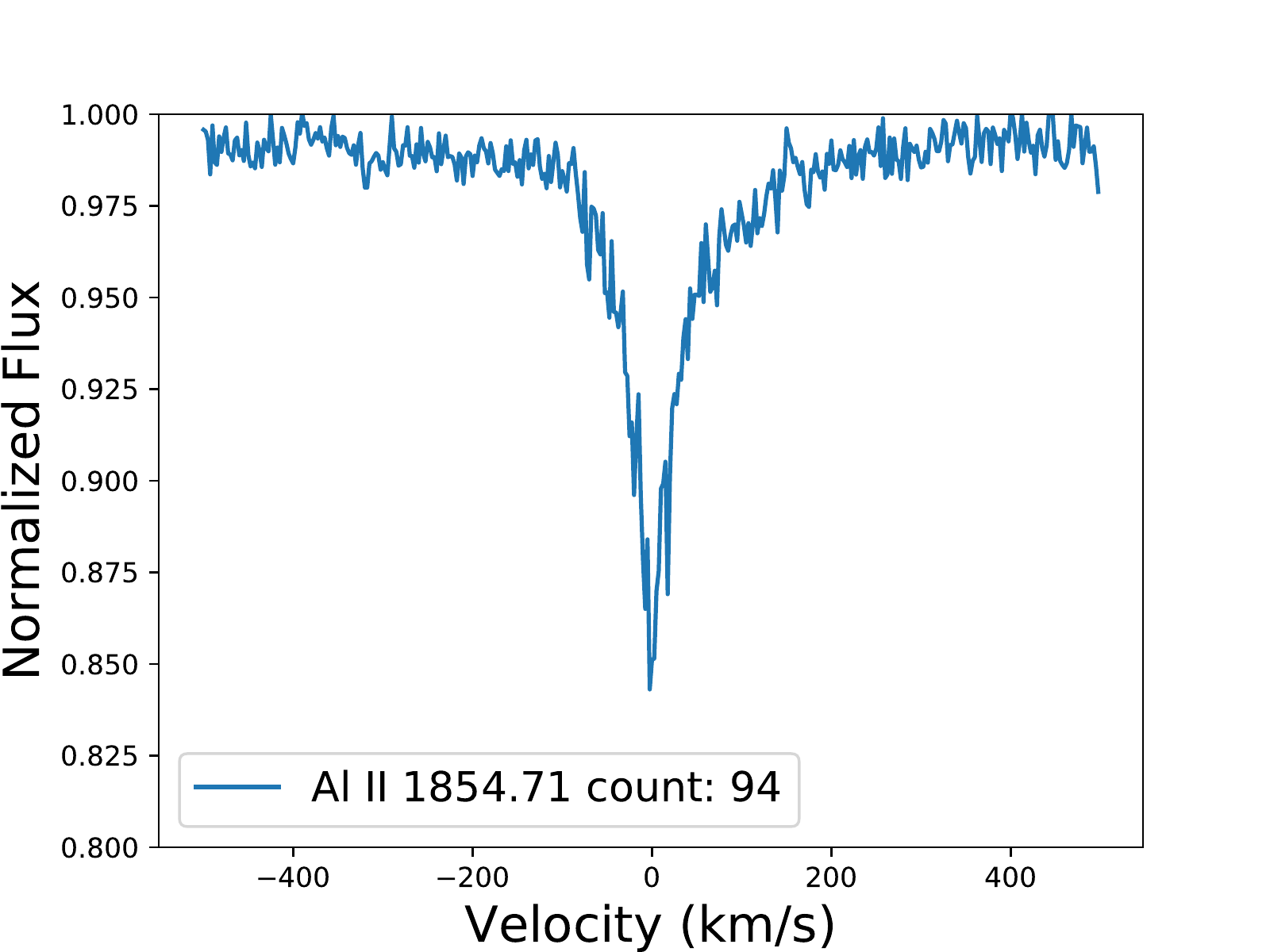}   
  \includegraphics[width=\columnwidth]{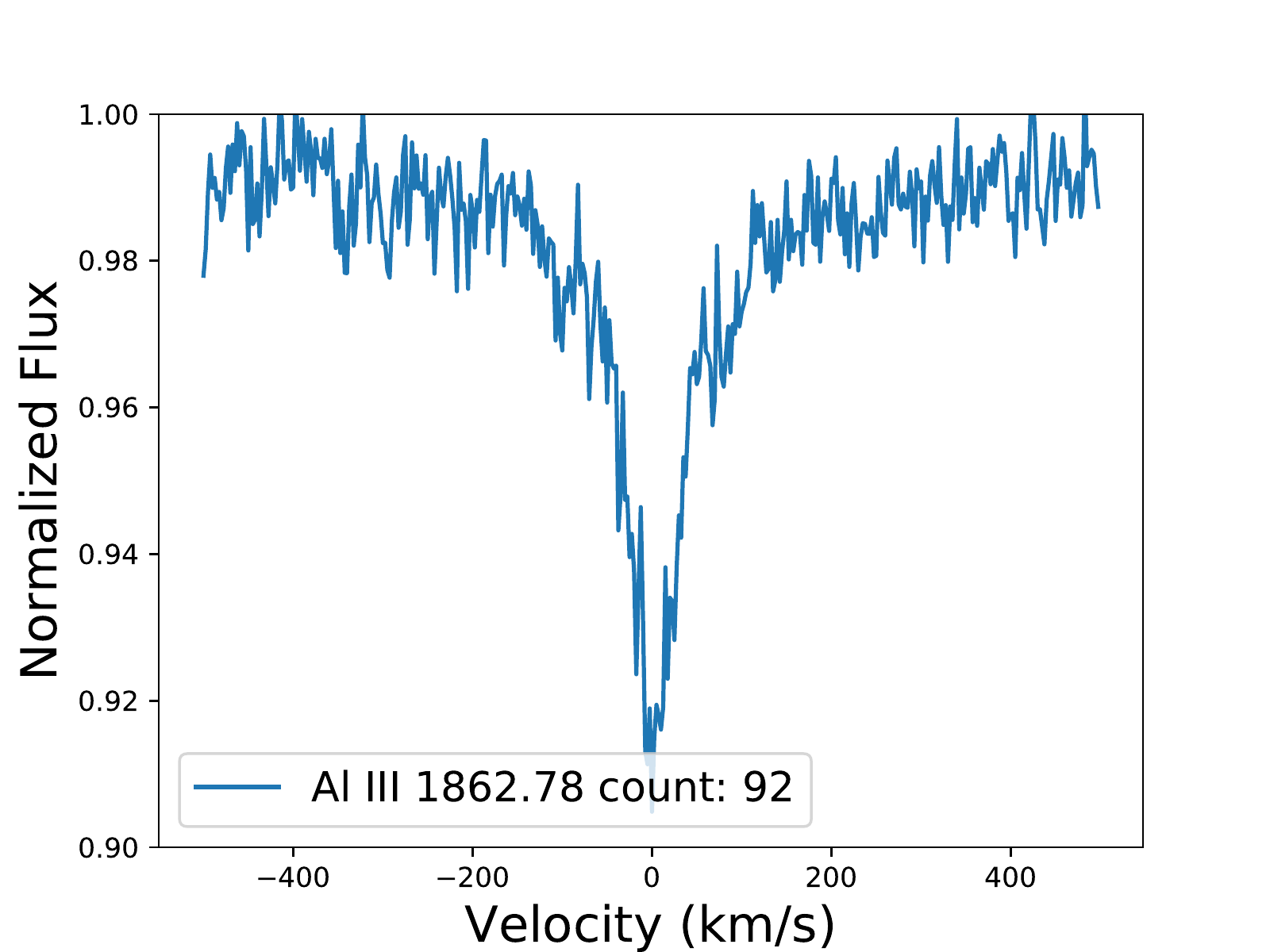}
   \includegraphics[width=\columnwidth]{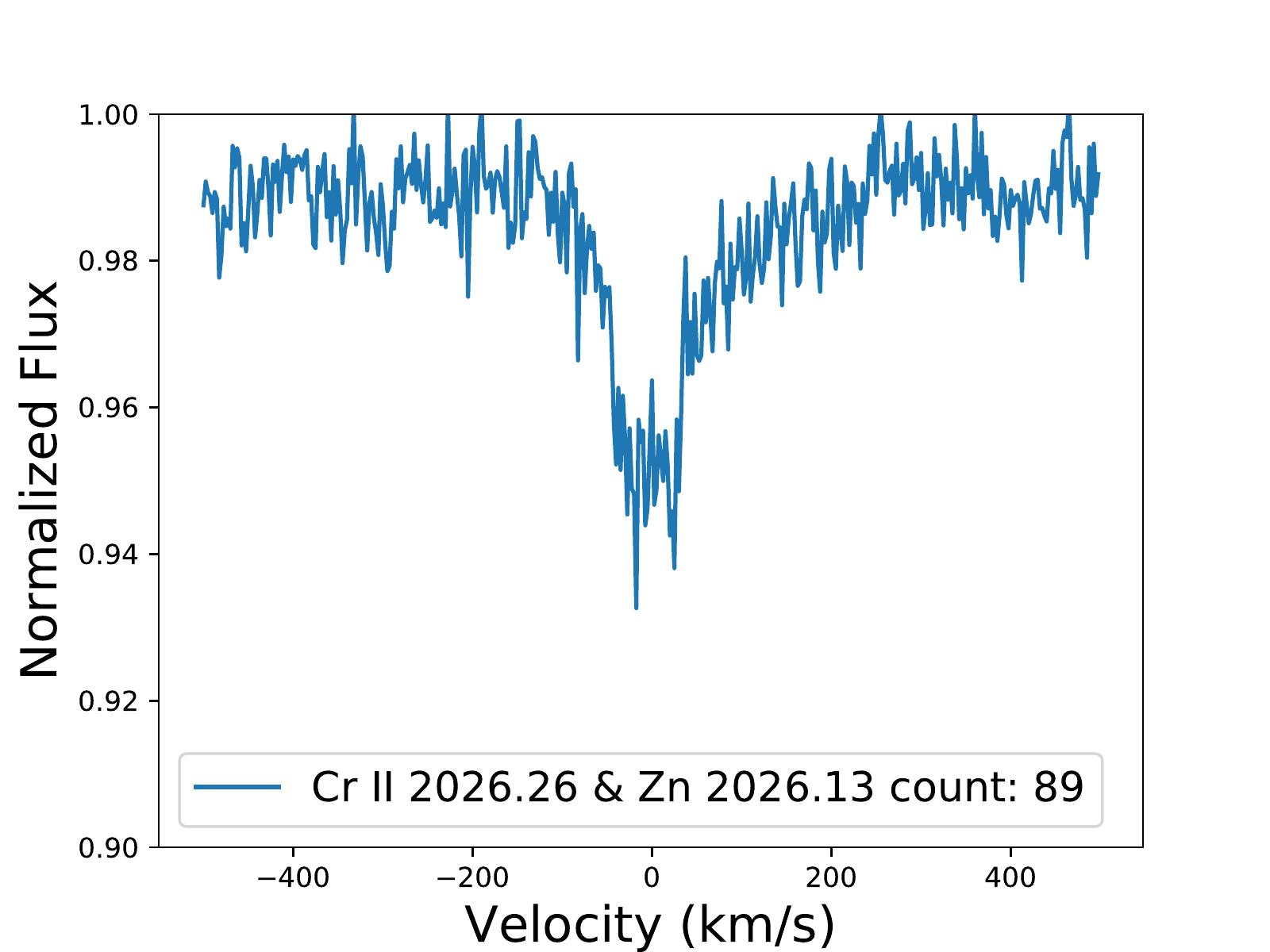} 
   \includegraphics[width=\columnwidth]{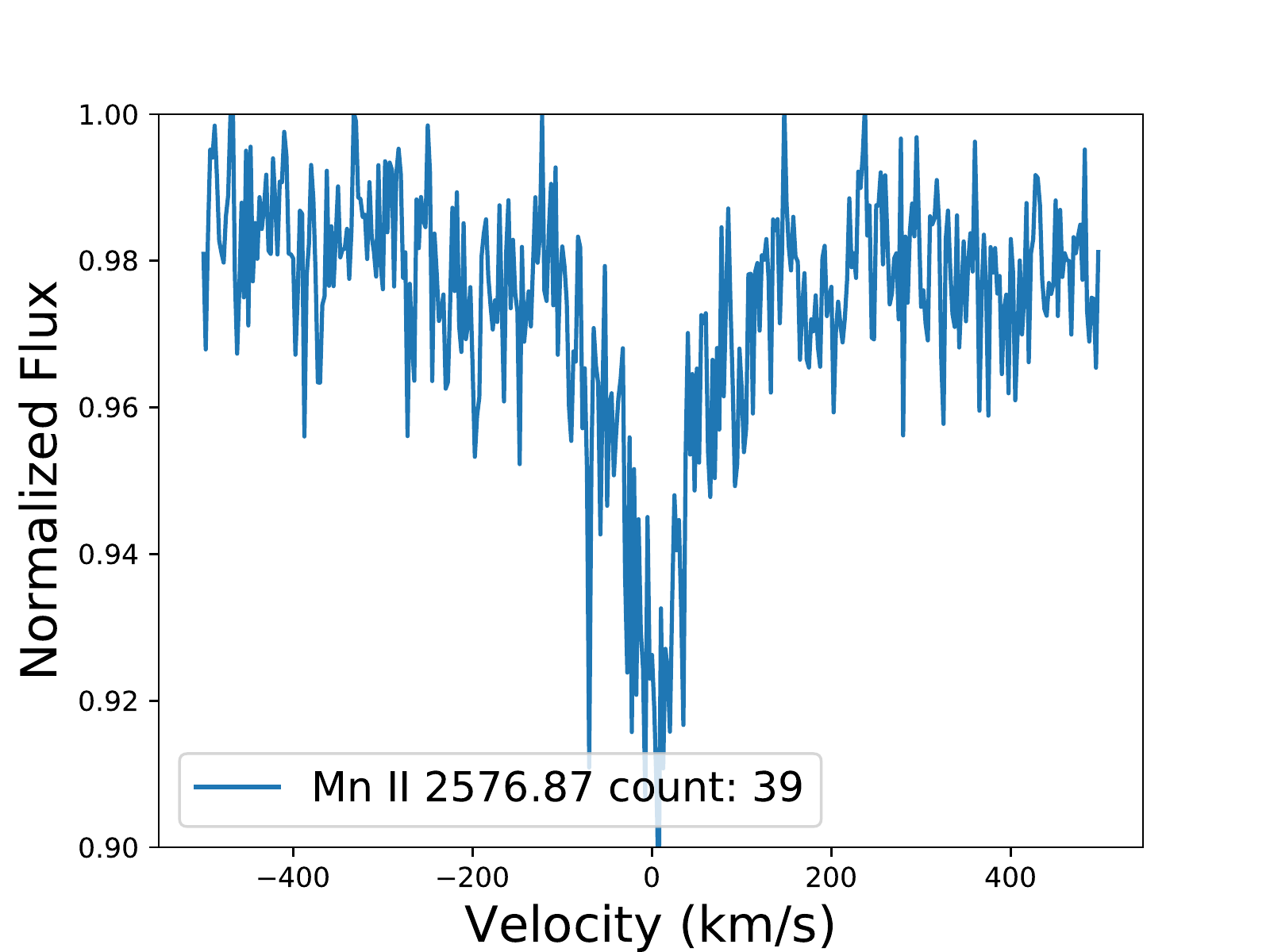} 
    \includegraphics[width=\columnwidth]{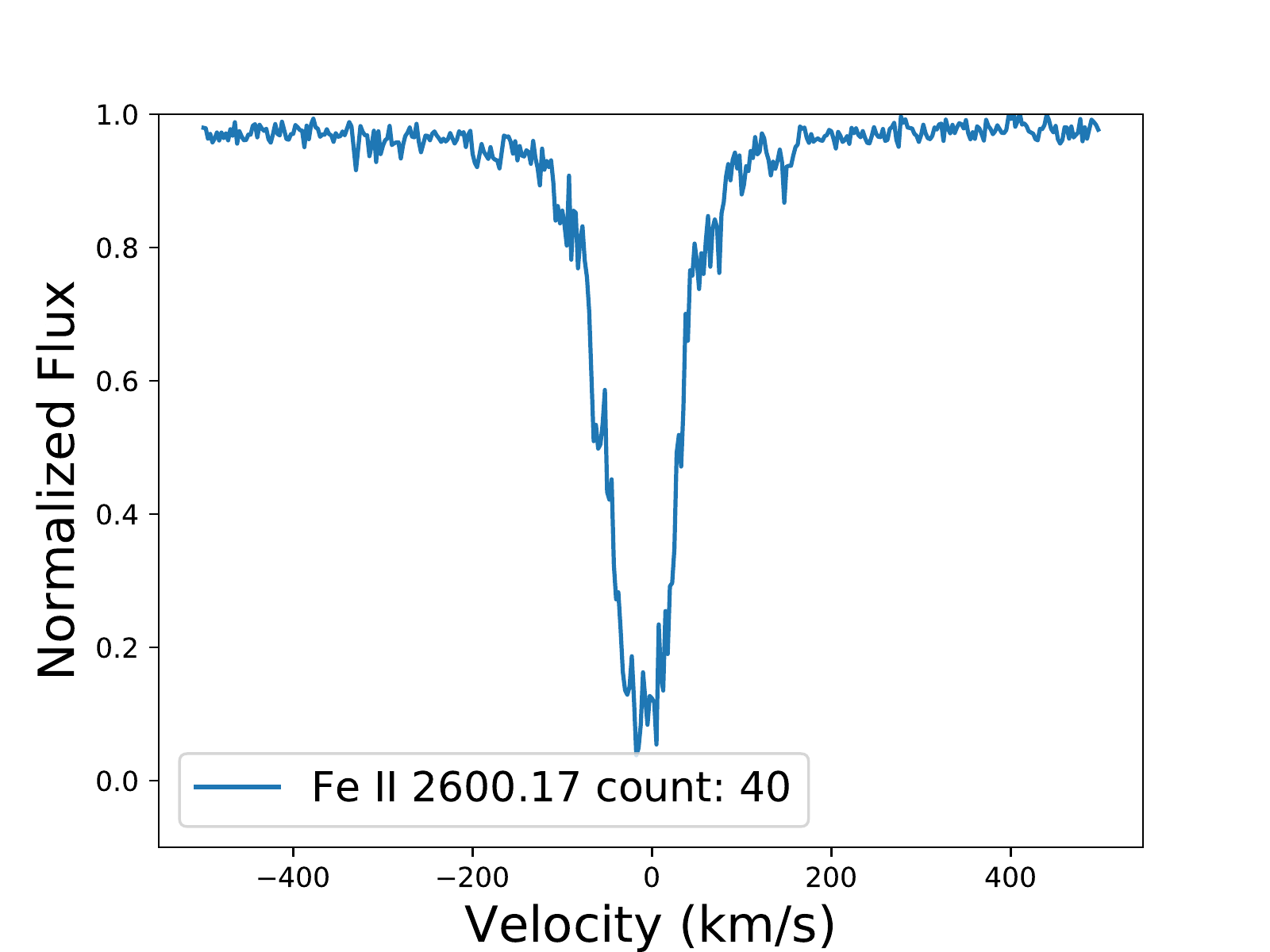}   \\

\end{figure*}

\begin{figure*}
\includegraphics[width=\columnwidth]{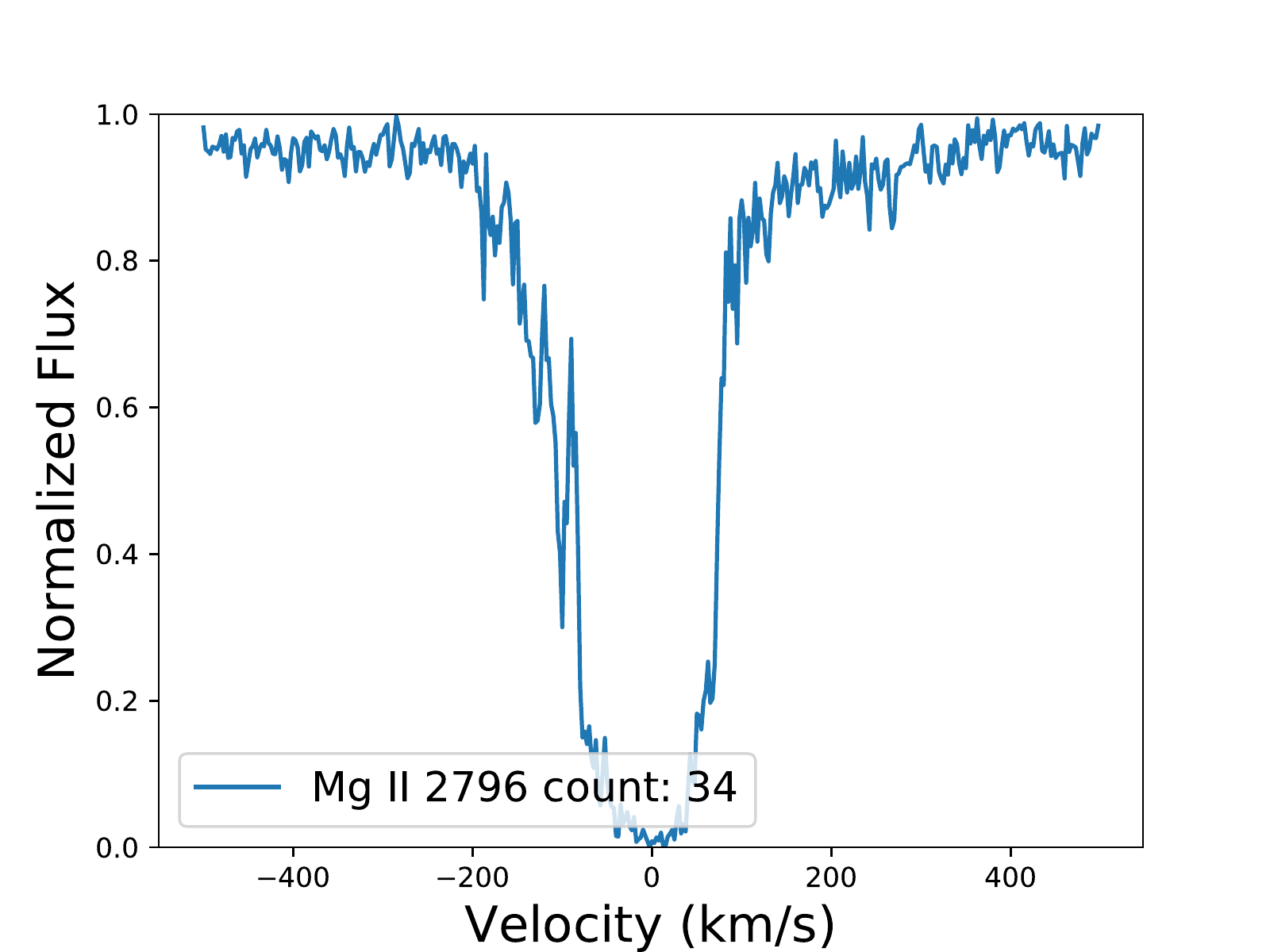} 
   \includegraphics[width=\columnwidth]{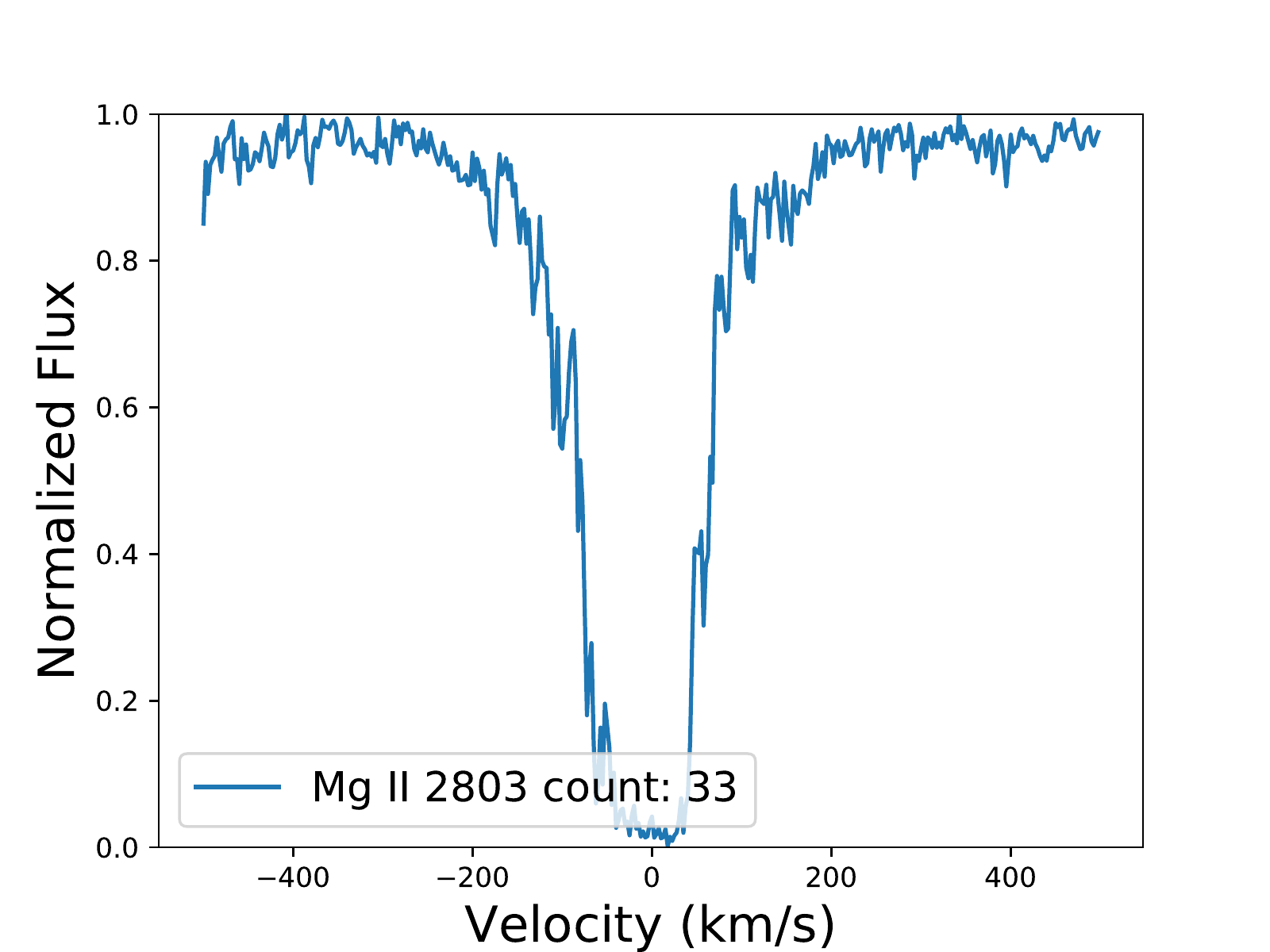} 
\caption{Stack of strong metal lines at the DLA redshift. The number of stacked normalised UVES quasar spectra are indicated in the legend for each element as well as the rest wavelength of the metal line. The mean redshift of the DLAs sample is $\mathrm{z_{DLA}= 2.5}$. The dropping continuum levels in the C\,IV panels indicate the present of the other member of the doublet, while in O\,I panel it is due to nearby SiII\,1304\,\AA\ absorption. }
\label{fig:StrongLinesI}
\end{figure*}
\label{lastpage}
\end{document}